# High Resolution Imaging in the Visible with Faint Reference Stars on Large Ground-Based Telescopes.

Craig Mackay[1]

*1 Institute of Astronomy, University of Cambridge, Cambridge CB3 0HA, UK*

**Abstract.**

Astronomers working with faint targets will benefit greatly from improved image quality on current and planned ground-based telescopes. At present, most adaptive optic systems are targeted at the highest resolution with bright guide stars. We demonstrate a significantly new approach to measuring low-order wavefront errors by using a pupil-plane curvature wavefront sensor design. By making low order wavefront corrections we can deliver significant improvements in image resolution in the visible on telescopes in the 2.5m – 8.2m range on good astronomical sites. As a minimum the angular resolution will be improved by a factor of 2.5-3 under any reasonable conditions and, with further correction and image selection, even sharper images may be obtained routinely. We re-examine many of the assumptions about what may be achieved with faint reference stars to achieve this performance. We show how our new design of curvature wavefront sensor combined with wavefront fitting routines based on radon transforms allow this performance to be achieved routinely. Simulations over a wide range of conditions match the performance already achieved in runs with earlier versions of the hardware described. Reference stars significantly fainter than I ~ 17m may be used routinely to produce images with a near diffraction limited core and halo much smaller than that delivered by natural seeing.

.

## 1. Introduction

Galileo recognised more than 400 years ago that the sharpness of the images he saw with his telescope was affected by atmospheric disturbances. Astronomers understand the importance of image sharpness delivered by a telescope systems. Sharper images reduce confusion between light from the target and that from other close by sources. They allow more of the light from a target object to be concentrated enabling photometry or spectroscopy to achieve a higher signal-to-noise ratio against the light from the night sky.

We now know that the image quality is largely set by the phase variance of the light across the aperture of the telescope. There are many techniques of minimising that variance including careful telescope design and siting. One of the most effective ways is to put a telescope in space. It is nearly 30 years since the launch of the Hubble Space Telescope (HST), an instrument which transformed our view of the universe by providing a tenfold step change in the angular resolution of visible and near infrared astronomical images. Although larger ground-based telescopes have a diffraction limit much smaller than the ~0.1 arcsec images routinely delivered by HST, progress in achieving better resolution has been slow, expensive and of rather mixed success. The degradation in image quality caused by the effects of atmospheric turbulence has been more difficult to overcome than anticipated.

Many major observatories have invested heavily in adaptive optic systems and good high-resolution images have been obtained particularly in the near infrared where the effects of turbulence are much less severe. In the visible, however, it is still the case that no adaptive optic system has achieved HST resolution on an HST size (2.5 m diameter) telescope on targets fainter than I~15. This lack of success inevitably calls into question the ambitious plans to correct the turbulent wavefronts entering the next generation of very large telescopes.

The majority of instruments in use have been designed to give a high order of wavefront correction on large telescopes using Shack-Hartmann wavefront sensors (see Section 6). These work by forming an array of separated images of a reference star from well-defined areas of the telescope pupil. The movement of each of these images as they are deflected by corrugations in the wavefront entering the telescope is tracked. This allows the system to compute the compensating deflections needed to correct the wavefront. Those corrections are fed to a fast deformable mirror so that the wavefront corrugations are substantially eliminated. The number of sub-images used (typically several per square metre of telescope aperture) and the readout rate to be used (typically several hundred Hz frame rate) constrain the brightness of the reference star. This is usually limited around I < 12-13, a level that ensures there is only a of very small chance of finding such a bright reference star close to a random science object. More recently, laser guide stars are being developed to improve sky coverage. In practice they create their own problems. For example, the laser beam does not pass through the same atmospheric turbulence as the science beam and produces images that are not at infinity. Even neglecting this focus anisoplanatism (also known as the cone effect) and other aberrations, laser guide images are seldom smaller than about one arcsec in diameter. They cannot generally be centred to better than one tenth of that size and so are unlikely to deliver an angular resolution significantly better than that of HST. That appears consistent with reported results so far. Results of recent tests of different LGS systems on Gemini South may be found in Marin (2018).

It is of great importance to astronomers to be able to work in the visible part of the spectrum since that is where much of our knowledge of the physical universe has been gained. It is in the visible (~ 300-1000nm wavelength) that we are best able to measure stellar types and metallicities, and where we know most about the astrophysics of hot gases in nebulae. The legacy of the Hubble Space Telescope has been its capacity to deliver images nearly 10 times sharper than most ground-based telescopes deliver on faint targets. There are too many science programs that have benefited from this to list here.

Wavefront correction will always be much harder in the visible than in the near infrared since the phase variance scales as $\sim \lambda^{-1.2}$ (Fried, 1967). However the rewards in terms of enhanced resolution of working in the visible rather than the infrared are considerable. An ideal system would indeed work in the visible on current 4-10 m class telescopes to deliver 20-50 milliarcsec angular resolution images that could feed efficiently a high-resolution spectrograph system. Such a system should be able to work with faint natural guide targets so that sky coverage approaches 100%, and it must deliver an isoplanatic patch size that allows reliable measurements of the characteristics and shapes of both guide and science targets over a significant field of view.

This paper will concentrate on strategies for minimising the effects of atmospheric turbulence in the visible. The technologies we use require fast full frame readout with photon shot noise limited performance. At present that limits us to silicon-based detectors such as CCDs and CMOS arrays which have a long wavelength limit in the region of 1000 nm. The resolution problem becomes much more tractable as one goes further into the infrared. This is one of the reasons why infrared performance of adaptive optics (AO) systems is emphasised. However we should not forget that any technologies that can be demonstrated in the visible on 5-10 m class telescopes in I band will work just as well in K band on 30 m class telescopes, something likely to be important for future extremely large telescope projects. Such developments will depend on the availability of fast framing detectors. Early versions are already in use and promise substantial improvements in the near future. The performance requirements are less demanding because, when working in the IR, photon rates are generally much higher than in the visible.

In recent years there has been a divergence in AO system development. One branch concentrates on the development of AO systems aiming to produce the highest resolution images for programs such as the detection of exoplanets around bright stars, often near naked-eye brightness (Sivaramakrishnan, 2001). The other strand which has had much less attention in recent years is targeted towards improving the resolution on much fainter images such as distant galaxies (Baldwin et al, 2008). Telescopes can now image remarkably faint objects yet the demands of astronomers to observe such objects at high resolution are not currently easily fulfilled.

## 2. Improving Resolution With Lucky Imaging.

Some progress has been made in delivering higher resolution from the ground but it is rather patchy. Some of the work has focused on Lucky Imaging (LI) either on its own on 2.5 m class telescopes or in combination with low order AO on larger telescopes. Simple LI is now widely used with over 300 research papers already published.

Originally analysed by Fried (1978) following earlier work by Hufnagel (1966), LI relies on taking a large number of images with exposure times short enough to freeze image motion due to atmospheric turbulence. A reference star in the field allows the relative sharpness of each image to be determined. The sharpest images are selected then shifted and added to provide a high resolution composite image. On a good site under typical seeing conditions it is possible to approach HST resolution routinely on 2.5 m class telescopes (Fig. 1). These results simply used LI techniques without any AO to support it.

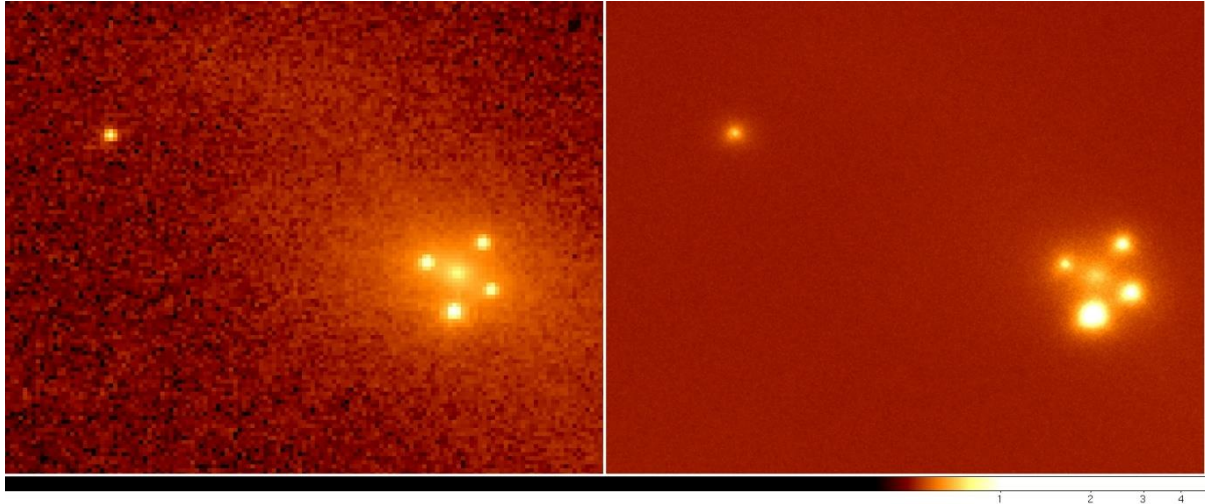

**Fig 1**: Images of QSO 2237+0305 (the Einstein Cross) gravitational lens. The light from a distant quasar is bent by a massive object in the core of a nearby Zwicky galaxy seen as the fuzzy object between four quasar images. The image on the left was taken with the HST Advanced Camera for Surveys (ACS) while the one on the right was taken by an LI camera on the 2.5m NOT telescope on La Palma, and selecting 30% of the images. Microlensing within the lensing galaxy causes the relative brightness of the four images to change on relatively short timescales hence the difference in flux ratios (from Mackay et al., 2012)

The statistics of turbulence makes the technique difficult to apply directly to use on larger telescopes as the probability of getting a sharp image is greatly reduced. The critical technology that made LI possible 30 years after it was first suggested by Fried was the development of high efficiency electron multiplying CCDs that could be read out at high frame rates with essentially zero read noise.

LI techniques have been described in detail by Baldwin, Warner & Mackay (2008) and references therein, as they apply to an HST size telescope in I band. The probability of recording a high resolution image under these circumstances is typically 10-30%. With larger telescopes the probability diminishes in proportional to the area of the primary mirror. This is because the number of turbulent cells increases as the area of the telescope mirror increases. The chance of a lucky exposure is thereby dramatically reduced. The power spectrum of atmospheric turbulence approximately follows the predictions of Kolmogorov (Tatarski, 1961) and is dominated by turbulent power on the largest scales. LI works best when the average number of turbulent cells across the diameter of the telescope is in the range of 6-18 (Hecquet & Coupinot, 1984, and Baldwin, Warner & Mackay, 2008). The only way that LI might work on a larger diameter telescope is if a significant additional part of the turbulent power can be removed so that the turbulent cell sizes are effectively increased, thereby compensating for the increased mirror diameter.

Noll (1976) has examined the contribution of atmospheric turbulence when expressed in terms of Zernike coefficients. He showed that eliminating tip-tilt errors reduces the turbulent power to only 13.3% of its original levels. In addition, he showed that removing an additional three Zernike modes after tip-tilt correction (Z4-Z6) leaves 6.9%. Further, removing a further four modes (up to Z10) leaves only 1.6% of the original power. It is clear from this that the considerable effort needed to remove even higher powers will only return a relatively modest improvement in image sharpness when 98.4% of the original turbulent power has already been corrected for.

In order to test whether the idea of removing turbulence beyond the simple tip-tilt scales removed by LI, a series of experiments were carried out in 2007 using a standard LI system placed behind the Palomar Adaptive Optics system PALMAO (Law et al, 2008). Conventional LI would be predicted simply not to work on a telescope of this size, giving a negligible fraction of sharp images even under the best seeing conditions. The PALMAO instrument was a relatively low order AO system with 241 active actuators and a 16 x 16 sub aperture Shack-Hartmann wavefront sensor. The results obtained were extremely encouraging (Fig. 2 and Fig. 3).

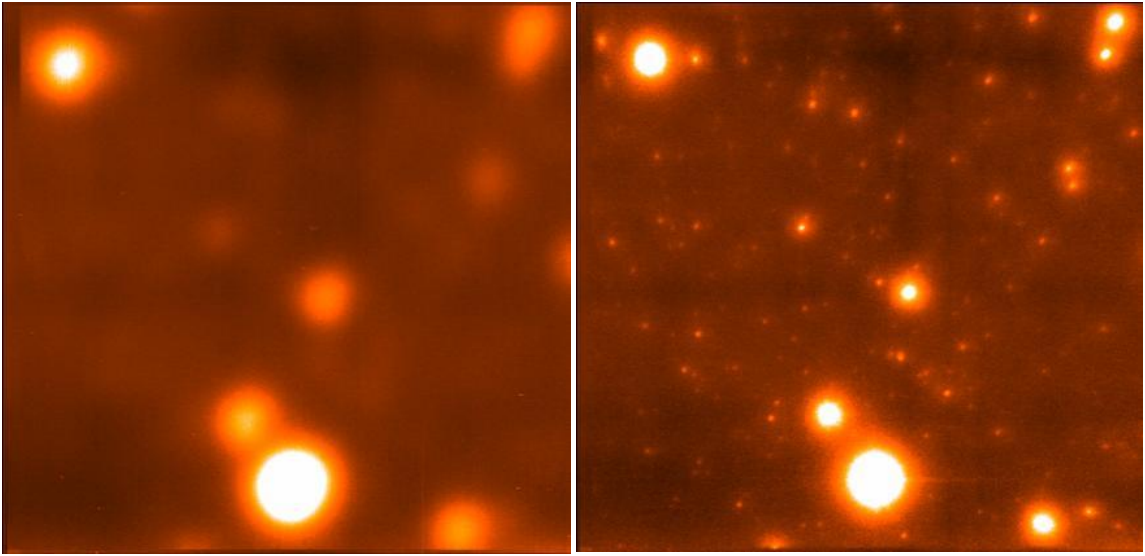

**Fig 2**. The core of the globular cluster M 13 in I band observed on the Palomar 5-m telescope. The left-hand image is with natural ~0.65 arcsec) seeing, and on the right hand is an image taken with the Lucky Imaging Camera behind the low order PALMAO adaptive optic system on the Palomar 5-m telescope. The resolution here is about 35 milliarcsec or about 3 times that of the Hubble Space Telescope. The total field of view shown is about 10 x 10 arcsec. This is the highest resolution image of faint objects ever taken in the visible or infrared anywhere from space or from the ground. The isoplanatic patch size is clearly large, much greater than 10 arcsec (from Mackay et al, 2012). The right hand image comes from a 10% LI selection.

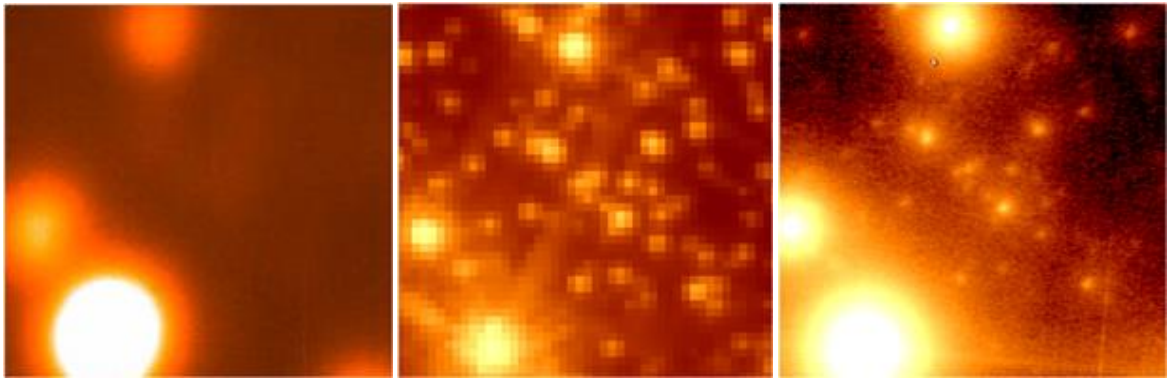

Fig 3. Comparison images of the core of the globular cluster M 13. The left-hand image is with natural (~0.65 arcsec) seeing on the Palomar 5-m telescope, then the Hubble Advanced Camera for Surveys with ~120 milliarcsec resolution (middle) and the Lucky Camera plus Low-Order AO image with 35 milliarcsec resolution (right hand) (from Mackay, et al., 2012).

Images with a limiting resolution of about 35 milliarcsec were obtained in SDSS-I band, centred on 771nm, and a bandpass of 91nm. These are the highest resolution images ever taken of faint targets in the visible or near infrared anywhere. It was clear that a similar instrument on an even larger telescope would undoubtedly have produced images with yet high angular resolution in the visible. The principal limitation of the PALMAO + LI instrument was that it needed the usual relatively bright reference object that is normally essential with Shack-Hartmann wavefront sensors.

These results showed that there was considerable potential in combining LI with low order AO. However these results showed that in order to work over a much larger part of the sky it was important to develop a new approach to sensing wavefront errors that allows much fainter reference objects to be used routinely. The technique had to be realisable with current technologies, both hardware and software in order to improve wavefront sensor performance dramatically and at the lowest photon levels.

This paper will suggest a rather different approach to achieve this. The architecture of most current AO systems originated nearly 50 years ago. Since then there have been radical changes in technologies that may be applied to these problems. We will explore how they can be used profitably to produce images approaching the diffraction limit of large telescopes using natural guide targets in the visible.

**3. Pragmatic Higher Resolution Imaging.**

An important concept in any practical development programme is the concept of "good enough". Engineers striving to produce the highest performance AO systems will often talk about the Strehl ratio achieved by the system. This is the ratio of the peak intensity of the restored image to the intensity it would have in the absence of any atmospheric turbulence using the same optical system. Often it is simply better resolution that is needed in order to allow nearby objects to be separated or to improve signal-to-noise in sky limited observations. Astronomers always ask for better performance but it is important to appreciate just what image quality is actually needed for a specific scientific application. Diffraction limited performance is simply not needed in every case. Any significant improvement is helpful but we must never lose site of the fact that, for the scientific programme in question, "good enough" is often a more realistic and achievable goal.

We can see this by looking at some of the science to be carried out with these systems. A good example here is the search for extrasolar planets which is often principally limited by scattered light from the star around which planets are being sought. In this case the highest Strehl ratio possible is critical to the success of the work. However for many other studies the ultimate Strehl ratio is much less important. It is the ability to resolve objects with confidence and to observe them with a well-defined and quantifiable point spread function even if that point spread function does not correspond to a particularly high Strehl ratio. The consistency of point spread function is much more important since that may be managed with deconvolution methods, for example, to establish meaningfully whether objects are resolved and what is their proper structure. In practice, images with a Strehl ratio as low as 0.1 are perceived visually as being sharp. Much higher Strehl ratios produce images that are perceived visually as being only marginally sharper. Yet it is the wish for these higher Strehl ratios that substantially drives a choice of technologies that are so difficult to implement with fainter science targets.

Our approach will be to look carefully at what we know about atmospheric turbulence and the extent to which it limits the performance of an imaging system. We will then look at the limits imposed by the sky in terms of the available target star brightness and surface densities as a function of galactic latitude. Following that we will consider new concepts of wavefront sensors that will allow us to work at significantly fainter levels than otherwise possible so that we can correct the turbulent wavefront in an efficient and timely manner.

**4. Atmospheric Turbulence Characteristics**

Atmospheric turbulence distorts the wavefront entering the telescope by introducing significant phase errors that smear out the final image. The turbulence that affects astronomical imaging is generated by interactions between different layers of airflow at a wide range of altitudes from the ground layer up to well in excess of 10 km. The stirring motion than many imagine dominates this turbulence in fact stops within a few turning times. We can see that by looking at high altitude contrails left by high-flying aircraft. At their altitude, also typically in the region of 10 km, complex patterns are generated rapidly that can persist for many tens of minutes. These represent regions of slightly different temperature that become frozen into the airflow. The same is true for astronomical turbulence: the turbulent patterns across a telescope entrance aperture are largely frozen into the airflow and the timescales over which the turbulence changes is comparable to the wind crossing time of the telescope.

In order to make some sense of the way the simulations are run and how the results should be interpreted we need to start by looking at the way that Zernike polynomials are used to analyse the generated wavefront. Zernike polynomials are two-dimensional functions which are added in different proportions to give the best representation of a single two-dimensional function which, in this case, represents the phase errors we wish to measure. We restrict the number of Zernike polynomials to represent any particular simulated wavefront and it is the wavefront constructed from those polynomials alone which is subtracted from the recorded wavefront to give the effect of correction on that wavefront.

Provided the signal-to-noise is high enough in the simulated wavefront then we can use increasing numbers of Zernike polynomials to provide a better and better approximation to describing the phase errors of the wavefront. However we will see that the amount of turbulent power that can be attributed to higher-order Zernike terms is very small so we expect diminishing returns as we increase the number of polynomials we use and particularly as we decrease the level of light in each simulation.

A representation of these polynomials is shown in Fig 4 (image from Wikipedia). The top polynomial represents the piston component and is ignored as it has no effect on the image quality. The next row ($Z_1$) represents simple tip-tilt while the following row ($Z_2$) accounts for the defocus and basic astigmatism terms. Each subsequent row represents progressively higher and more complicated structures within the wavefront. What we need to do is to find out for each photon rate per frame how many Zernike terms can we afford to use in order to deliver the best results of higher Strehl ratio combined with narrower halo profiles.

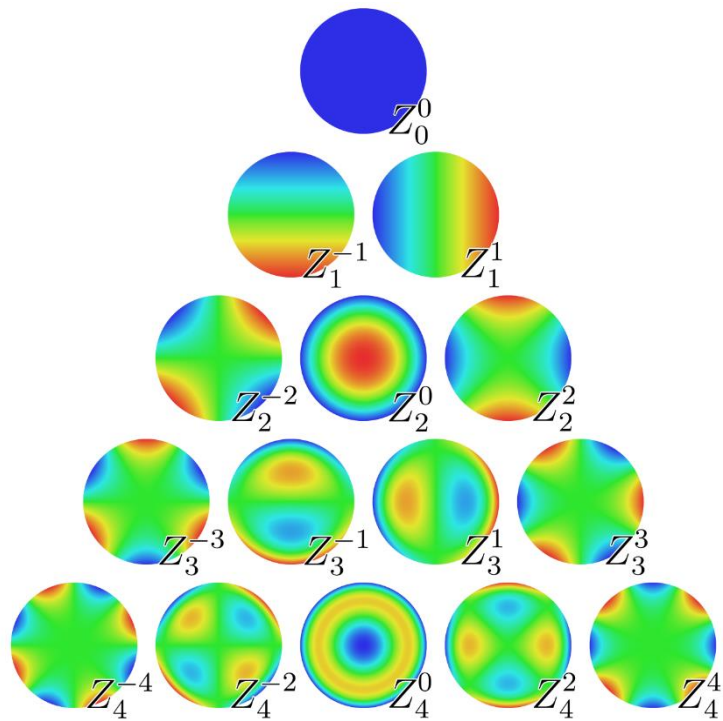

**Fig 4**.. This shows the appearance of the first 15 Zernike modes that are used to analyse and reconstruct the wavefront patterns used in this paper. The top line shows simply the piston error which we ignore. The next two are simply tip tilt. As we use the increasing number of modes we see that the structure within them gets smaller and smaller. We will only be able to correct for the highest modes if there is an adequate signal-to-noise within the typical scale size that corresponds to those modes. Image from Wikipedia.

The power spectrum of atmospheric turbulence approximately follows the predictions of Kolmogorov (Tatarski, 1961) and is dominated by turbulent power on the largest scales. Noll (1976) has examined the power spectrum in atmospheric turbulence when broken down into individual Zernike coefficients. These results are shown in Table 1. It is clear that the power in individual Zernike terms falls very rapidly. Indeed once the first six Zernike terms are corrected the residual is very small indeed. Measuring and then correcting higher order terms will require a wavefront sensor signal level much greater than needed for the lower order terms. For example, in order to remove Z11 and Z12, each of which only contributes two parts per thousand of the total turbulence power will require a signal for the wavefront sensor of millions of photons per de-correlation time.

This is particularly important when developing AO systems intended to work on relatively faint targets. Trying to fit and correct an excessive number of Zernike terms will simply add noise to the fitted wavefront and that in turn will degrade the corrected image unless the reference star is relatively much brighter.

| Zernike terms removed | Residual power fraction | Residual Power per term |
|---|---|---|
| 1 | 1.000 | - |
| 2 | 0.565 | 0.435 |
| 3 | 0.130 | 0.435 |
| 4 | 0.108 | 0.022 |
| 5 | 0.085 | 0.022 |
| 6 | 0.063 | 0.022 |
| 7 | 0.057 | 0.006 |
| 8 | 0.051 | 0.006 |
| 9 | 0.045 | 0.006 |
| 10 | 0.039 | 0.006 |
| 11 | 0.037 | 0.002 |
| 12 | 0.034 | 0.002 |

**Table 1.** (from Noll, 1976): distribution of turbulent power between Zernike terms.

## 5. The Challenge Of The Sky

Bright reference objects are very scarce, particularly at high Galactic latitudes. Counts of stars have been presented by Simons (1995) in R-band. Using a mean colour index of R-I of 1.5 mags, stars with I =20.5 arc found at around one per square arc minute at high Galactic latitudes, I =19.3 at 60°, and I =18 at 40°. When considering reference objects at these magnitudes, it is important to remember that many faint galaxies have compact cores which, if small enough, could also be used as reference objects. At I =20, galaxies are more common than stars at the Galactic poles (Pozzetti et al, 1998).

Following Kaiser et al. (2000) we expect that a good back illuminated CCD will accumulate about one detected electron per second from an object with an I magnitude given by

$M_I = 23.5 + 5 \log (\text{Diameter in m})$

Our simulations are calculated in terms of the number of photons detected per frame time. What this implies will be become clear later in the paper when the detailed optical configuration of our instruments is described. When working in low order the phase patterns are close to being achromatic simply because it changes relatively slowly with wavelength ($\sim \lambda^{-1.2}$) and therefore a significantly wider passband may be used for the wavefront sensor to increase the photon rates while still allowing the science beam to be more narrowly filtered. In addition, the latest deep-depletion CCDs deliver a very much greater quantum efficiency throughout the red end of the spectrum and indeed the sensitivity extends now well beyond one micron wavelength. If we assume that we use all the light longward of 660nm (the short wave limit of I-band) in conjunction with a deep depleted CCD then the photon rates detected could be nearly a factor of 10 higher than for a standard back illuminated CCD in I band. This in turn means that the magnitudes quoted in the figures may be increased by 2.5 magnitudes. Most importantly they bring the sky surface density of reference objects to a level of several per square arc minute. We will see later that the isoplanatic patch size is typically greater than one arc minute diameter therefore we may be confident of finding reference objects almost anywhere on the sky.

## 6. Wavefront Sensor Technologies.

### 6.1 Shack-Hartmann Sensors.

Light from a distant star produces a flat wavefront before passing through the atmosphere. Refractive index variations in the atmosphere produce instantaneous distortions of this wavefront which introduce curvature on all scales. In order to measure and correct this curvature we need a wavefront sensor. For many years the wavefront sensor technology of choice has been the Shack-Hartmann sensor. Light from a bright guide star is fed to an array of lenslets placed in the pupil plane of the telescope. Each lenslet images the guide star, thereby giving an array of images on a regular grid in the absence of atmospheric turbulence. Wavefront distortions produce phase gradients across the aperture of each lenslet leading to the displacement of each sub image. By measuring the relative positions of each sub image it is possible to derive the curvature in the wavefront across the pupil plane (Fig 5).

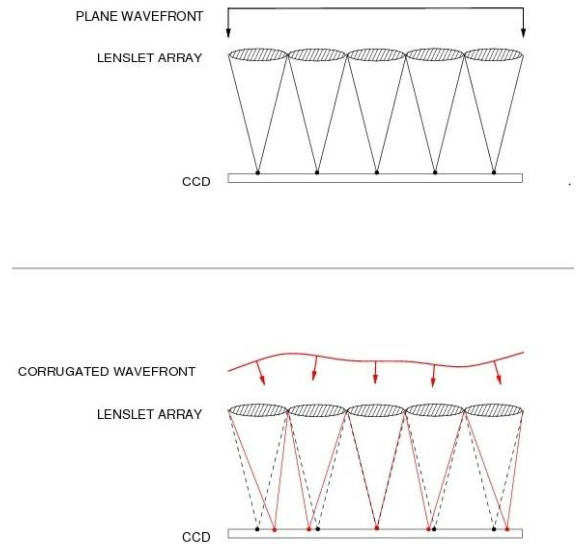

**Fig 5:** The principle of the Shack-Hartmann wavefront sensor. A distorted incoming wavefront enters the sensor. Part of it is then passed through an array of lenslets located in a pupil plane. The centroid position from each lenslet image is measured relative to the position recorded for an incoming flat wavefront. By comparing the offsets, the slope of the wavefront at this position may be calculated and then combined with the other values to determine the incoming wavefront distortion across the entire pupil (Murphy, 1992). Image courtesy of Prof Vik Dhillon, Univ. of Sheffield, UK.

The Shack-Hartmann sensor became popular because in the 1970s only CCDs operated in slow scan mode could achieve very low read-out noise at millisecond frame rates and then only with a substantial degree of charge binning in both directions (Mackay, 1986). This allowed performance relatively close to that set by the photon signals in each sub image. Unfortunately the regular lens spacing set the resolution in the pupil plane used to find the wavefront curvature. It was very inflexible and difficult to reconfigure. Most importantly, however, was that by spreading the light from the guide star amongst many hundreds of sub-images, bright guide stars became essential if each sub-image was to achieve the signal level of about 100 photons per lenslet per frame, a level necessary to get good positions of each sub image.

In practice this meant that guide stars generally have to be brighter than perhaps I ~13, a level at which the surface density of guide stars on the sky is extremely low. The isoplanatic patch, the angular scale over which the wavefront distortions were reasonably close to those detected with the guide star, was rather small, typically only a few arcsec in diameter (Sarazin, M., & Tokovinin, A., 2001). There are other issues with Shack-Hartmann sensors. The frame rate has to be set to match the wind crossing time of each lenslet aperture. This requirement can force detectors to achieve very fast read-out rates of several hundred Hz. This in turn increases the required brightness of the guide star as enough photons per aperture must be gathered per frame time. An additional overhead with Shack-Hartmann sensors is that the array of sub-image displacements measured from each frame require a matrix inversion to convert into wavefront curvature data used to drive a deformable mirror. The inversion can be computationally difficult particularly as the overall system latency between detecting the wavefront and correcting for the errors must be minimised.

It should be clear from this that our desire to work with very much fainter guide stars is incompatible with using a Shack-Hartmann sensor with many hundreds of elements and running at several hundred frames per second. If we are to make progress towards reducing the effects of turbulence on our images then we need to take a very different approach.

**6.2. Curvature Sensors.**

A different approach to wavefront sensor technology is to try to measure the wavefront curvature directly. Most of the curvature wavefront sensors used in astronomy examine defocused images of the reference object by adjusting the focus of the telescope on either side of the nominal position. Analysing the shape and sharpness of these out of focus images allows the curvature of the input wavefront to be determined. Curvature wavefront sensors have been adopted by astronomers relatively slowly because of difficulties particularly with the detectors and the format needed to be efficient. However they generally use many fewer detector elements and that leads to significant sensitivity advantages. Racine (2006) has looked at the limiting sensitivities of wavefront sensors deployed at that time on telescopes and found that the limiting sensitivity of curvature sensors (Roddier, 1988) are typically 2.5 magnitudes fainter than Shack-Hartmann sensors although this difference is progressively reduced with higher order correction requirements. One of the best-known and most successful curvature sensors was Hokupa'a, a natural guide star, curvature-sensing AO system built by the University of Hawaii (Graves et al, 1998) with a 36 element deformable bimorph mirror. The flexible mirror allowed slightly out of focus images to be recorded on either side of the nominal telescope focal plane. It was driven with a loudspeaker voice coil to move the sensing plane between the two focal positions.

The implementation of the curvature sensor concept was very demanding when that work was done. Although by modern standards this is a rather low order AO sensor it was remarkably successful in delivering significant image resolution enhancement. Recent developments, particularly in detector technology, allow a much simpler and much more efficient approach. We have seen earlier, however, that low order AO corrections are often all that is needed and indeed we will see that that is often all that is possible should we have to use much fainter guide stars.

There are a number of designs of curvature wavefront sensors which mostly use avalanche photodiodes (APD's) as light detectors (Baldwin et al., 1996). These are single element detectors that function rather like a photomultiplier. The image plane is segmented in a complicated pattern with each segment fed to a separate APD. It was the availability of these APD's that helped in allowing curvature sensors to be developed successfully.

Since then there have been major strides in the development of essentially noiseless two-dimensional imaging detectors. The first was the electron multiplying CCD (EMCCD), developed by E2V (Chelmsford) (Mackay et al, 2004). Under normal operation at high frame rate the read noise of a conventional CCD is prohibitively high, precluding its use on faint astronomical targets. In EMCCD's, however, the output register of the device is extended with one gate in three clocked with a high enough voltage to trigger electron multiplication as the charge is transferred along. Total gains with ~500 elements and a gain per stage in the region of 1-2% can produce an aggregate gain of hundreds or thousands. This is much higher than the read noise of the output amplifier and therefore the equivalent read noise can be very much less than one electron RMS even with pixel rates as high as 30 MHz (Mackay et al., 2012a)

There are now CMOS detectors which have all the advantages of the best astronomical CCD detectors in terms of high quantum efficiency, low dark current and cosmetic quality. The latest devices are now being made with sub-electron read noise even when working at very high frame rates that can be in excess of 1 kHz even at full resolution.

We know that current focal plane curvature sensors offer significant sensitivity gain over Shack-Hartmann systems. The many advances in technology make it important to reconsider the design principles of wavefront curvature sensors and this is what we have done. We have concluded that the best way forward particularly when working with faint reference stars and therefore relatively low order correction is to carry out the curvature sensing in the pupil plane and not in the focal plane of the telescope.

# 7. Pupil Plane Curvature Sensors and Simulations.

## 7.1. Pupil Plane Curvature Sensors

Our approach is not to image the focal plane but work close to the telescope pupil plane. The principles of such a sensor have been described by Guyon (2010). The propagation of light through the telescope pupil can be understood as follows: in the entrance aperture (the pupil plane) of the telescope the illumination is uniform although the phase errors that affect our images caused by atmospheric turbulence are naturally present. If we image not in the pupil plane but in planes displaced from the pupil plane we find that the uniform illumination breaks up into patches. Comparing images taken on either side of the pupil we see that those parts of the wavefront that are converging will change from darker to brighter as they propagate through the pupil. Similarly wavefront phase curvature in the opposite sense will cause a patch to become darker as it propagates through the pupil. We can detect these near-pupil images and analyse them to extract the wavefront curvature. An example of how the light propagates through the pupil plane is shown in Fig 6. This figure considers simply the light that actually passes through the entrance aperture of the telescope and what the intensity distribution of that light would be over a range of distances from the pupil. In order to see a significant intensity structure in this light propagation distances in hundreds of kilometres must be considered.

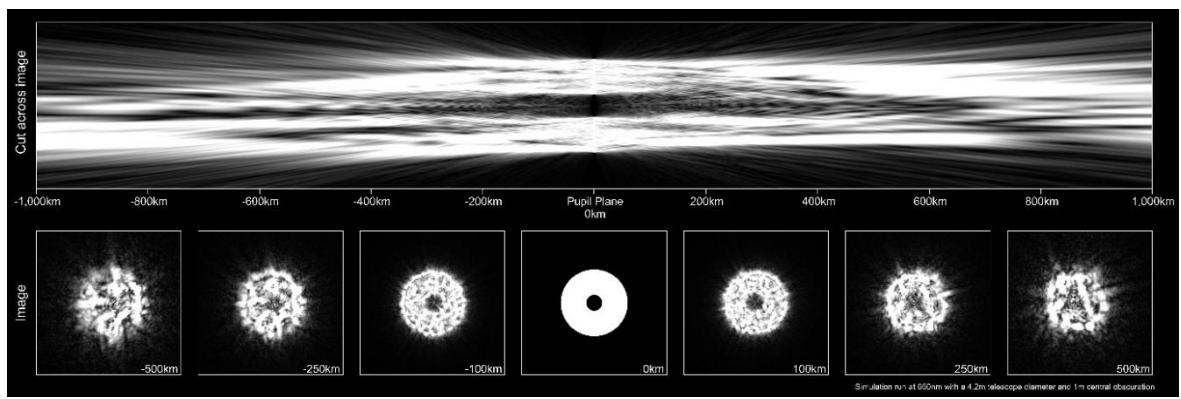

**Fig 6.** The propagation of light through the pupil of the telescope. In the pupil (the entrance aperture of the telescope) the illumination is uniform (central image above). We then examine how the light that actually passes through the pupil would have looked before it entered and after it left the pupil. On either side of the pupil the image breaks into a speckle-like pattern. As we image further and further from the pupil plane the speckle pattern is increasingly dominated by large-scale structures. Note that the distances shown on this figure have nothing whatever to do with the altitude at which turbulence is generated. They simply indicate how the propagation at different distances provides particular sensitivity to particular scales of turbulent power. The wavefront propagation is substantially achromatic in low order allowing the use of broad response bands with a curvature sensor (Guyon, 2010). Figure from Mackay et al. (2012b).

Rather than mechanically move a single detector to image alternately on either side of the pupil plane we find it more efficient and convenient to split the light so that both planes are imaged simultaneously. This means that the pair of out-of-pupil images and therefore the curvature is precisely measured at one instant. This allows the frame rate to be chosen and optimised for the timescales of the wavefront curvature evolution. By using a pair of EMCCDs working in synchronism we can be certain that a pair of precisely complementary images have always been taken.

Processing these images gives the wavefront curvature in the pupil plane (see section 9.3). The deformable mirror is then adjusted to correct the curvature errors in a reimaged pupil plane. In practice, this turns out to be much more straightforward than with the Shack-Hartmann sensor mainly because the processing yields the curvature directly and above all quickly. It is essential when using any turbulence correction procedure that the correction is derived and applied before it changes significantly. In order to investigate the effectiveness of these procedures further we carried out a series of simulations to improve our understanding of the optimum optical arrangement for such a sensor and the performance that might be achieved with different degrees of Zernike correction, different photon rates and different average seeing.

We are fortunate in having a significant amount of real experience of attempts to undertake high resolution imaging on a range of telescopes all on good sites (although often the weather was much less than "good"). This includes several LI-only (no AO) runs on the 2.5m NOT telescope on La Palma, the 4.2 m WHT telescope also on La Palma, the 3.6m NTT telescope on La Silla in Chile. We also obtained many valuable results with LI combined with low order AO using the Palomar 5 m telescope in conjunction with the PALMAO low order adaptive optic system. These datasets have allowed us to compare different aspects of our simulations in some detail with the results actually obtained at real telescopes.

**7.2. Pupil Plane Curvature Sensor Simulations.**

We wanted to carry out realistic and detailed simulations to discover just how well it was possible to restore the distorted wavefronts caused by atmospheric turbulence if it was measured, quantified and corrected for promptly. There are several components needed to complete this.

- We need to look carefully at the way that we simulate the phase patterns experienced on different sizes of ground-based telescopes under a variety of seeing conditions.
- We need to identify detector subsystems capable of working at fast frame rates yet at the lowest photon rates possible. The system needs to be highly efficient from a DQE point of view while still maintaining full imaging integrity. The detector must work with the readout noise low enough to permit operation at very low photon rates.
- We need to develop a procedure for reconstructing wavefront curvature quickly and efficiently. It is particularly important that the technique should be stable under very low light conditions when individual frames may be momentarily starved of light. The technique must also be able to manage poor seeing and retain good stability under all conditions.
- We need to examine just how many Zernike terms need to be corrected for in practice and therefore what the necessary photon rates must be to achieve the performance we want, under different seeing conditions.

We will always be conscious of the need to identify what is "good enough" for the particular scientific programme. We do not seek perfection! With any form of LI we are always able to look at an individual frame and decide not to include it in the summed totals if it is not good enough for our purposes. This is important because the seeing can change on very short timescales indeed (Baldwin et al. 2008), making a degree of image selection highly desirable.

These simulations were intended for instruments already in the design phase and so we only assume currently available technology.

**7.3. Model Turbulent Screen Generation.**

In order to simulate the behaviour of low order adaptive optic systems when combined with LI systems we have identified the importance of the largest scales of atmospheric turbulent power. A common and frequently used approach to generating theoretical turbulent screens makes an assumption of periodic boundary conditions. Such periodic phase screens are unphysical as they do not have a significant average slope and therefore do not model the tip tilt component of the wavefront which is a fundamental part of the turbulent spectrum. A summary of this problem and how to address it can be found in Johnston and Lane (2000).

A more detailed account of the technique we have used to generate a model of the Kolmogorov phase fluctuations over an aperture is described by Johnston (2000). This uses the midpoint displacement algorithm as described by Harding et al. (1999). Using this method we have subjected the screens produced to a variety of tests that demonstrate that they do indeed provide the correct Kolmogorov spectrum of turbulence predicted by the work of Noll(1976) including tip-tilt. We used this method to produce much larger phase screens than we needed so that we could simulate the effects of wind motion across the aperture of the telescope. It also allowed us to study the decorrelation of restored point spread functions as a function of angular distance from the reference star used and therefore to determine the effects of anisoplanatism.

**7.4. Example Instrumental Configuration.**

In order to understand how a pupil-plane curvature system might be employed in a real instrument we show the outline of one of our designs in Fig 7. AOLI (Adaptive Optics Lucky Imager) was developed in Cambridge for use on the 4.2 m William Herschel Telescope (WHT) on La Palma.

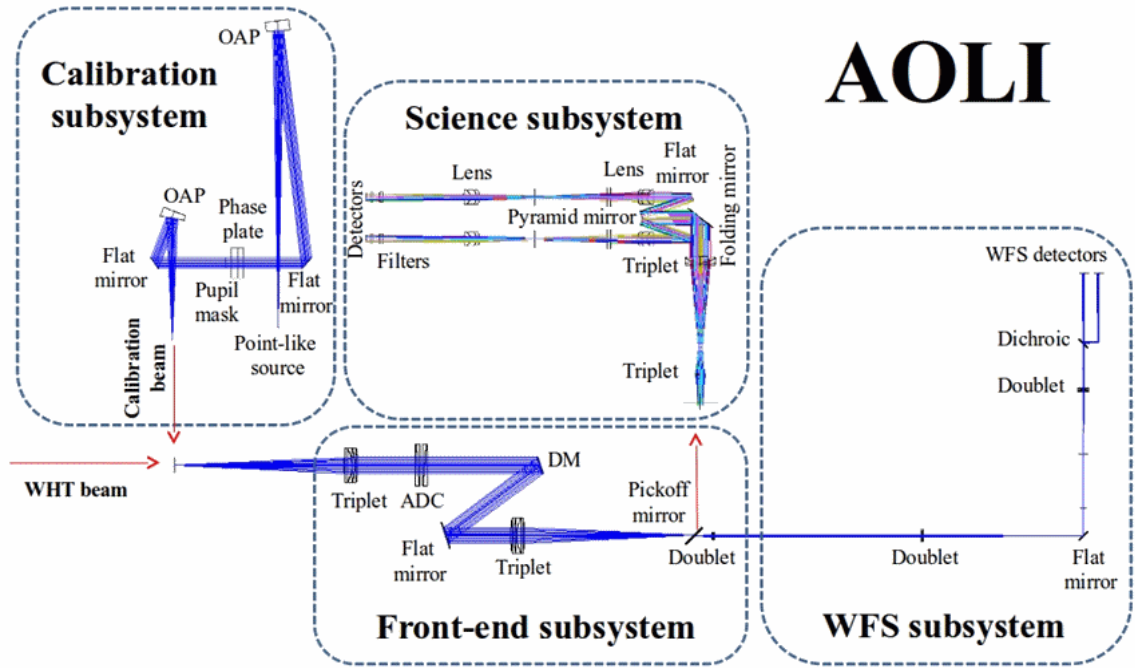

**Fig 7**: the outline schematic of AOLI (Adaptive Optics Lucky Imager). The beam comes from the telescope on the left of the diagram and can be replaced by a calibration beam to allow it to be tested fully in the laboratory. An optical relay focuses an image of the pupil plane on the deformable mirror in the front-end subsystem. A small hole in a pickoff mirror allows most of the field light to be deflected to the science subsystem while the light needed to sense the wavefront errors enters the WFS subsystem. A pair of doublets produces a greatly diminished image of the pupil only a few millimetres in diameter. Two copies of the wavefront are generated by a 50-50 beam splitter labelled on this diagram as a dichroic, with different path lengths so that the detected wavefront images correspond to either side of the pupil plane shown in Fig 6. (Crass, King & Mackay, 2014).

The design of AOLI therefore allows pairs of images to be collected simultaneously taken from close to the pupil plane. The optical configuration in the instrument reproduces the propagation distances in a greatly diminished scale so that the pupil is reduced in size to the region of 1-2 mm. The pair of near-pupil images are then sensed with a photon counting image detector. High-speed photon counting array detectors have been developed in recent years in the form of the electron multiplying CCD (EMCCD) (Mackay et al., 2012a). The choice of magnification is quite important with photon counting. In order to detect a single photon we must minimise the risk of two photons striking one pixel in one frame time. This limits the maximum photon counting rate in a detector to a fraction of one photon per pixel per frame time. Using a larger detector area becomes attractive because of that but the influence of clock-induced charges (CIC) can be a problem. These CIC events are already present in all standard CCD systems but are not usually detected because of the much higher equivalent background noise on those systems. With an EMCCD, these events look exactly like photons and are generated at a rate of $10^{-3}$ to $10^{-4}$ per pixel per frame. Using a detector area of 100 pixels diameter (about 7000 pixels area after correcting for the pixels covered by the central obscuration of a real telescope) this effect can give several events per frame if the CIC on the system in use is relatively high. Fortunately, careful electronic design allows the lower figure to be realised (Mackay et al., 2012a). We have found that working with the wavefront detector area of about 100 pixels in diameter is a reasonable compromise in practice.

The EMCCD generates events with a large dispersion in amplitude. This compromises the overall signal-to-noise ratio in a way equivalent to halving the detective quantum efficiency (DQE) of the detector. This reduction in DQE can be eliminated by thresholding to replace each pixel with a 0 or a 1. This must be done in real-time, before the phase calculations are carried out. Fortunately it is a very fast process with a modern computer system. Strategies for managing this are described by Basden et al. (2003).

The simulation package that we have created is based on the original work and the Matlab kernel from Marcos van Dam and collaborators (van Dam & Lane, 2002). It allows many of the important parameters that are used to be adjusted. That package was enhanced to allow modelling of the effects of wind speed, instrumental latency, detector and exposure time constraints. All the essential observational parameters could be varied (telescope aperture size, pre-and post-pupil propagation distance, isoplanatic patch size, seeing) allowing a wide range of parameters to be studied effectively.

These include:
- Telescope specific parameters: telescope diameter, resolution of the turbulent screen detected.
- Central wavelength (set here to 770 nm default, which is the central wavelength of the SDSS-I filter.
- Atmospheric seeing in the range 0.4-1.5 arc seconds corresponding to values of the Fried diameter $r_0$ in the range 0.45-0.12m, with most work around 0.6 arc second seeing, $r_0 \sim$ 30cm.
- Parameters affecting turbulent screen smearing and system latency including windspeed, response time for the wavefront processing, the response time for the deformable mirror and wavefront image readout, essentially detector frame rate.
- Photon rate per wavefront sensor frame. AOLI divides the light from the reference star into two which are imaged at the same distance from the pupil plane but on opposite sides. The photon rate given in each simulation is the rate per detector. This means that the true rate required is twice that listed.
- The package has the capacity to synthesise simulated images in positions offset from the reference star to examine the overall system isoplanatic patch size. The patch numbers and separations may be set.
- Propagation distance described above. The precise choice of distance here is not particularly critical and we have settled on 200 km. It is important to realise that this propagation distance has nothing whatever to do with the altitude of the turbulence that we might be trying to correct for. It is what we would detect were we able to take images at that distance on either side of the pupil plane, using only the light that actually propagates through the pupil of the telescope. By de-magnify the pupil diameter, the propagation distances are reduced by the square of the magnification factor. For a 4.2 m telescope and a 1 mm relayed beam, 200 km propagation distance is reduced to under 100 mm.
- Number of Zernike terms to be corrected for ranging from 3 (simple tip tilt) up to a maximum of 36. For computational simplicity we limited this number to be one of 3, 6, 10, 15, 21, 28, 36.
- The detectors recording the two separate near-pupil images must be precisely synchronised so that each image refers to exactly the same time interval.
- The images are flat-field corrected if necessary before being processed by the wavefront fitting software. The processing time required for that is included in the latency figures used.

**7.5. Wavefront Processing Procedures.**

We start with a high resolution turbulent screen which is then degraded in a variety of ways. Firstly we process the wavefront using only a limited range of Zernike terms to the number we want to measure and correct for in the wavefront. We next smooth the turbulent screen in one dimension as this minimises the effect of the motion of the screen caused by wind. It is then sample offset by an amount dependent on detector readout time, instrument processing time and deformable mirror latency.

The wavefront is then sampled and limited to the diameter of the telescope and propagated in opposite directions to the positions of each detector. This allows us to record the intensity at those distances from the pupil plane. Here the phase errors in the wavefront are increasingly manifest as intensity patterns in the recorded near-pupil images. We wish to explore particularly the possibilities of deriving the wavefront aberrations using images obtained from faint reference stars where the number of photons recorded per frame could be small. It is essential that whatever method we use is stable in those regions where there are few or no photons at all. Any suggestion of spurious errors would certainly cause serious problems. The recorded analogue images are then populated with the selected number of photons using Poisson statistics. All the photons are given the same intensity as would be generated by a true photon counting electron multiplying CCD camera system as we are using in our instruments.

Following the work of van Dam & Lane (2002) we have based our wavefront recovery algorithms on the use of Radon transforms. These have seldom been used in astronomical applications (though see Starck et al, 2003) but are very well studied and widely used in, for example, x-ray computerised tomography systems where they are particularly successful. The radon transform uses each recorded image projected along a series of vectors to give a sequence of one-dimensional integrated profiles. The number of vectors needed depends on the number of Zernike terms being recovered. We used eight projections for up to 10 Zernike terms, 16 vectors for 15 or 21 Zernike terms and 32 vectors if we were using 28-36 terms. As a projection in opposite directions of the same pattern is identical the number we have to compute is half that shown. The patterns from the two detected images are processed to produce a least-squares fit to the phase pattern in the pupil plane of the setup. This is the fit we use.

Each phase pattern processed in this way is used to generate an image of the reference star. Those images are selected in terms of their maximum peak brightness as is widely done with LI described above. We also make a small adjustment to the position of the brightness peak of the image. This is because any error in the fitting can produce a residual tip-tilt error which we can correct with standard LI shift and add methods. The processing procedures are mathematically relatively simple. They consist of matrix multiplications associated with the Radon transform, together with one-dimensional interpolations. Computationally these processes are very straightforward and with fairly standard PCs will contribute very little to the overall system latency in a real-life application.

The software developed for these simulations has a great number of variables. However in order to compare our results directly with existing datasets derived at telescopes on good observing sites we have restricted the range of variables significantly. All our simulations were carried out at 0.77 microns wavelength (SDSS I-band). After some experimenting to maximise the quality of the wavefront fit, we selected a fixed propagation distances of 200 km on either side of the pupil plane. We also fixed the wind speed at a standard 8 m/s (the median windspeed for the La Palma Observatory) and used a detector frame rate of 25 Hz. The effect of a significant wind speed is simulated by smoothing the phase screen in one dimension. We have experience of using the ALPAO DM241 deformable mirror for the AOLI instrument described above which has a fast response that adds very little to the overall system latency when running at 25 Hz.

Trials were run as sequences of 1000 independent frames. Although the software allows the simulation of a continuous phase screen moving across the aperture of the telescope we have simply recalculated the phase screen for each frame independently. We found we could work with fainter reference stars using our knowledge of the most recent phase reconstruction as a template for the next but have not included that here.

The simulation procedure starts by generating a phase screen that is considerably larger than the aperture of the telescope. This ensures that phase fluctuations outside the nominal diameter of the pupil are included correctly as they may influence the propagated beam. For each frame we recover the phase from the detected images at the photon rates we have selected. The difference frame between simulated and recovered is then used to generate a corrected frame from which we create a corrected image (Fig 8). We can see the success by looking at histograms of the variance of the wavefront phase in each of the frames before and after correction (Fig 9).

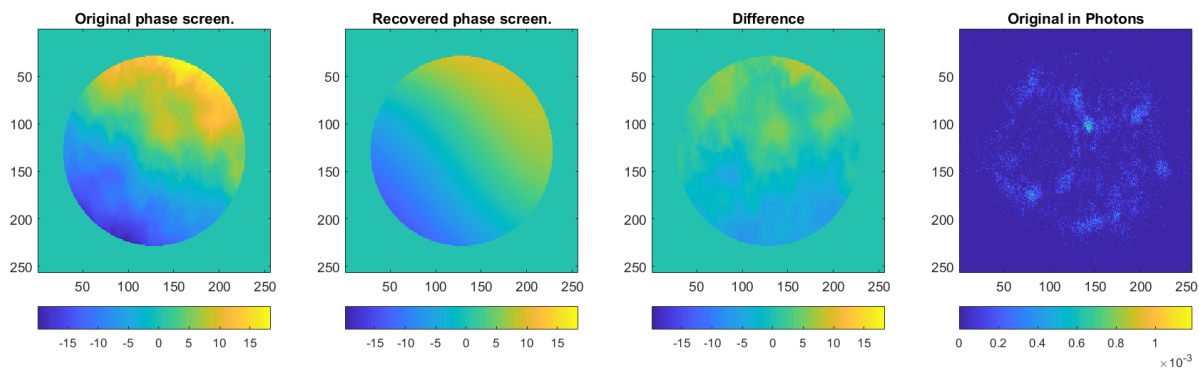

**Fig 8**: the sequence of images generated from a single frame. The original phase screen is shown on the left followed by the recovered phase screen. The third image shows the difference while the fourth shows the appearance of the original photon distribution across the pupil plane for this particular frame of the simulation. Simulation from 4.2m, 0.6 arcsec seeing, correcting 10 Zernike modes, high photons per frame.

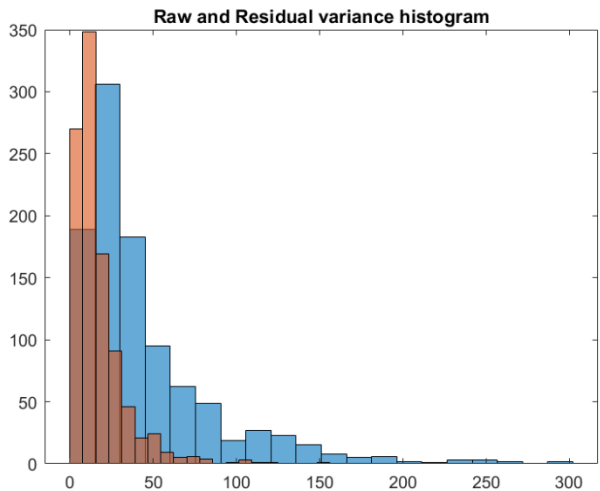

**Fig 9**: the improvement in the image quality is most easily seen by examining the histogram of all the images in a sequence of frames. We see here that the variance distribution for the blue bars (raw data as synthesised) are very much more extended than the orange bars for the pupil phase variance after correction. Simulation from 4.2m, 0.6 arcsec seeing, correcting 10 Zernike modes, 300 photons per frame level.

Once the sequence of 1000 frames had been processed to recover the simulated phases, lucky images were constructed using the best (judged by peak intensity level) using 1%, 5%, 10%, 20%, 30%, 50%, and 100% of the images in the sequence. Those images may be displayed as an array of images as shown in Fig 10 which also shows the Strehl ratio of each of the accumulated lucky images. The last sub- image shows the appearance of the summed raw data before any processing is carried out. This raw data image is what would be obtained with the conventional long exposure image. The Strehl ratio is the fraction of the peak brightness that would be found in the absence of any atmospheric turbulence.

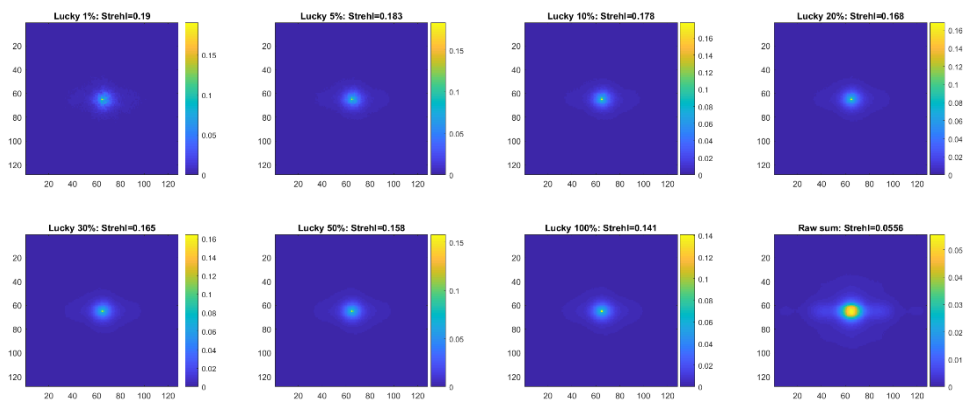

**Fig 10:** An array of 7 lucky images plus the simple sum of the raw image frames in a typical sequence of 1000 trials. In this case we see that the Strehl ratio even if 100% of the corrected images are used is improved by a factor of ~2.5. A more restrictive lucky selection further enhances the Strehl ratio. The smallest fraction of images taken from a sequence of 1000 frames involves a very small number and therefore the Strehl ratio displayed is likely to be affected significantly by small number statistics. Simulation from 4.2m, 0.6 arcsec seeing, correcting 10 Zernike modes, 300 photons per frame.

The strategy we used in running these simulations was to establish the degree of image improvement achievable with relatively bright guide stars and a range of correction limited to low order Zernike modes. We then expanded our simulations space to a range of reference star brightnesses, Zernike modes correction, different levels of atmospheric seeing, windspeed and detector frame rate, and finally telescope size. The ones that were studied particularly were relevant to the telescopes we had easiest access to, specifically the 4.2 m William Herschel telescope, 8.2 m GranteCan telescope and the Nordic Optical Telescope 2.5 m telescopes on the island of La Palma.

The theoretical point spread function predicted from Kolmogorov turbulence models (Fried, 1978) is known as a Moffat profile (Racine, 1996, and Trujillo et al, 2001 and references therein). This profile has a core that is essentially Gaussian but with significantly broader wings. In practice astronomers find the broad halo to be wider and more intense than predicted by the models. This is discussed by Racine (1996). The reasons for this are not fully understood but are probably a combination of light scattering in the atmosphere, from microscopic defects on the surface of mirrors and lenses, and internal reflection from optical components including back reflection from detector surfaces. For many astronomical applications the precise details of the actual image profile delivered by any telescope system needs to be understood if actual telescope images are to be interpreted correctly (Trujillo et al, 2001).

We combine LI selection techniques with low order AO correction. Under a wide range of conditions (telescope size and seeing quality) the star image profiles are those expected with a partially compensated Kolmogorov turbulence (Hardy 1998 and references therein). The images produced by our simulations consist of a compact central core close to being diffraction limited surrounded by a wide halo with the halo size smaller than the raw image. Increasing the degree of the AO correction has only a small effect on the width of the central core but a more significant one in reducing the size and intensity of the halo. The quality of the AO correction is improved with higher photon rates per frame and by correcting for more Zernike terms but only if the photon fluxes are high enough for that to work. When photon rates are low, attempting to correct for too many Zernike terms simply adds noise to the process and degrades the image profile.

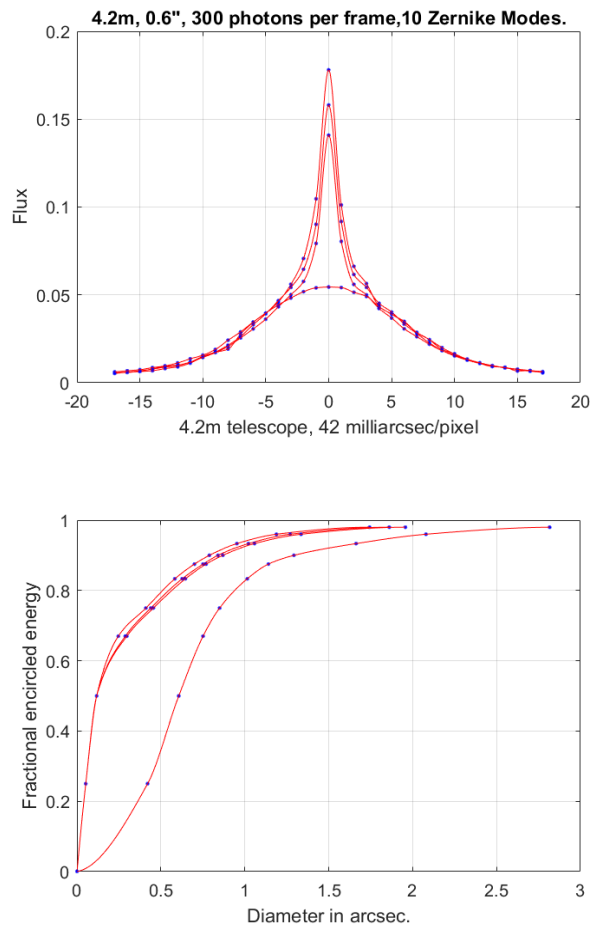

**Fig 11:** a typical set of profiles produced by the simulation package on a typical reference star. Simulation from 4.2m, 0.6 arcsec seeing, 300 photons/frame and correcting 10 Zernike modes,.
(a) The upper panel (a) shows one-dimensional profiles through a set of four images. The lowest curve in that frame shows the point spread function of the uncorrected telescope seeing. The three narrow profiles above show the image after correction for the given number of Zernike modes with 100%, 50% and 10% selection by image sharpness respectively. We see that the Strehl ratio even with 100% selection is improved by a factor of about 2.5 - 3 and the width of the central core is close the diffraction limit of the telescope in I-band.
(b) The lower panel (b) shows the fractional encircled energy profiles extending much further from the centre. Here the lower profile shows the raw data and the following three profiles the 100%, 50% and 10% selections respectively by image sharpness.

The output from each simulation run can be seen most easily in the form shown in Fig 11. In Fig 11 (a) we see 4 separate profiles. The lowest profile shows a one-dimensional cut through the directly summed reference star before any processing. This is the image that would be obtained in conventional astronomical imaging without any lucky imaging of any other attempt to improve the image resolution. The peak brightness (essentially the Strehl ratio of the image) is lower for larger telescopes and under conditions of poorer seeing. However under all conditions we find that the Strehl ratio for the images formed by correcting each one and then combining all the images shown by the second lowest profile in Fig 10 (a) shows a value larger than the raw data Strehl ratio by a factor of 2.5-3. The width of the central peak even with 100% selection is close to the diffraction limit of the telescope. Using more restrictive selections (50% for second top curve and 10% for the top curve) we see the Strehl ratio of each image is even higher. The image quality improvements continue as we are increasingly restrictive with the selection of only the best images (see Fig 10, for example). The diameter of the central core does not change much but increasingly the light in the surrounding halo is pulled in towards the centre producing significantly tighter images. We can see this more clearly in Fig 11 (b). Here the lower profile is that of the raw data while above are three profiles from the 100%, 50% and 10% images respectively.

We can understand how this works as follows. Without any adaptive optics, standard LI removes the tip-tilt component of the turbulence which makes up over 87% of the total power. This tip-tilt is what limits the size of the turbulent cell size defined as the typical radius of a cell with an RMS wavefront aberration across the cell of one radian. This gives a value of $r_0$ ~ 30 cm for the cell size with 0.6 arcsec seeing. With tip-tilt correction such a turbulent cell is effectively increased in size by a factor of 2.5-3 giving a turbulent cell size $r_0$ of only slightly less than 1m. If we then try to improve the correction by adding in compensation with low order adaptive optics we compensate for defocus and astigmatism by removing Zernike terms, numbers 4-6. This further increases the equivalent size of the turbulent cells and making the chance of getting a good sharp image even better. However the ability to remove progressively higher Zernike terms depends on the photon levels recorded in the frames from the wavefront sensor. Eventually the degree of image quality improvement reaches a maximum. Beyond that level we find that the image quality worsens because there are not enough photons to allow the higher Zernike modes to be established with an adequate signal-to-noise. That in turn leads to noise being added to the solution which in turn degrades the images.

## 7.6. Detailed Simulation Results.

We will discuss in much more detail later to interpret the levels of photons per frame using the simulation into a specific magnitude for a reference star. For the moment, however, we can use the conversion that for a 4.2 m telescope as described above running at 100 photons per frame with a frame rate of 25 Hz corresponds to a star with I~17.5.

We assume that the system is used with a genuine photon counting imaging system that produces counts with no dispersion in amplitude. We have used EMCCDs in photon counting mode (Basden et al, 2003) with event thresholding. Some of the simulations produce acceptable results with very low photon count rates. In practice the user must determine whether such rates are achievable when the background sky level may make that harder.

Each of the simulation runs produces data of the form shown in Figure 10. Rather than reproduce a large number of figures we extract the relevant data from the plots and display them in tabular form. We have found that the figures tend to disguise subtle but important differences between different simulations and so we prefer the use of these tables. It is important to remember that these simulations are statistical in nature and therefore there are likely to be small differences between values that ought to be identical. We have made no attempt to disguise that.

| 4m2 Diameter, 0.6" seeing, 10,000 Phot./frame, 10 Zernike Modes. | | | | |
|---|---|---|---|---|
| | | Image diam. | Image diam. | Image diam. |
| Selection | Strehl ratio | 25 % | 10 % | 4 % |
| 10% LI | 0.21 | 0.31 | 0.68 | 1.06 |
| 50% LI | 0.18 | 0.38 | 0.75 | 1.14 |
| 100% LI | 0.16 | 0.39 | 0.77 | 1.21 |
| Raw data | 0.057 | 0.84 | 1.28 | 2.06 |

**Table 2**. This table shows the results of a simulation, here for a 4.2 m telescope with 0.6 arc second seeing, photon rate of 10,000 photons per frame and correcting for the first 10 Zernike modes. The table then gives the peak Strehl ratio for each of the selections and the corresponding width of the profile at the level 25 % encircled energy, 10 % and 4 % of the peak central brightness.

The left-hand column shows the percentage of images selected in that row. The first number shows the Strehl ratio for that selection and then the diameters of the images at points which are well down towards the outer parts of the profile and set at levels of 25%, 10% and 4% of the peak brightness. The numbers indicated are the diameters in arc seconds.

We see therefore in Table 2 that the Strehl ratio is highest with the smallest selection (10% lucky imaging) and that the diameter of the profile at the 25% point is 0.31 arc seconds, 0.68 arc seconds at the 10% point and 1.06 arc seconds at the 4% point. Again, with larger selection fractions the image diameter values increase significantly although at every selection level they are markedly better than that achieved with just the raw data without any processing.

**7.7.    4.2 m Telescope Results.**

To better understand the performance of a 4.2 m telescope with low order adaptive optics combined with Lucky Imaging we ran a series of simulations starting with a high photons per frame number of 10,000 and examining the output images using between 3 and 15 Zernike terms. Using only 3 Zernike terms corresponds to conventional lucky imaging, as those terms only account for the tip-tilt part of the turbulent profile.

The results in Table 3 show that the Strehl ratio of the simulated images gradually improves with the use of more Zernike terms. The Strehl ratio for 10% selection is significantly better than it is for 50% and 100%. The Strehl ratio increases gradually as we go from 3 to 10 Zernike terms corrected. However the size of the halo around the central peak also improves up to (and we find beyond) 15 Zernike terms, at every part of the profile of the halo.

| 4m2 Diameter, 0.6" seeing, 10,000 Photons per frame, 3/6/10/15 Zernike Modes. | | | | | | | | | | | | | | | | |
|---|---|---|---|---|---|---|---|---|---|---|---|---|---|---|---|---|
| | Strehl ratio | | | | Image diam. in arcsec, at 25 % | | | | Image diam. in arcsec at 10 % | | | | Image diam. in arcsec, at 4 % | | | |
| Selection | NZ 3 | NZ6 | NZ10 | NZ15 | NZ 3 | NZ6 | NZ10 | NZ15 | NZ3 | NZ6 | NZ10 | NZ15 | NZ 3 | NZ6 | NZ10 | NZ15 |
| 10% LI | 0.19 | 0.20 | 0.21 | 0.23 | 0.35 | 0.36 | 0.31 | 0.29 | 0.75 | 0.69 | 0.69 | 0.63 | 1.14 | 1.11 | 1.08 | 1.02 |
| 50% LI | 0.16 | 0.17 | 0.18 | 0.20 | 0.42 | 0.40 | 0.37 | 0.32 | 0.83 | 0.72 | 0.74 | 0.67 | 1.24 | 1.20 | 1.14 | 1.06 |
| 100% LI | 0.14 | 0.15 | 0.16 | 0.18 | 0.46 | 0.44 | 0.38 | 0.34 | 0.88 | 0.74 | 0.76 | 0.70 | 1.33 | 1.26 | 1.21 | 1.12 |
| Raw | 0.057 | 0.057 | 0.057 | 0.057 | 0.84 | 0.85 | 0.85 | 0.84 | 1.28 | 1.23 | 1.30 | 1.31 | 2.07 | 2.09 | 2.09 | 2.07 |

**Table 3.** This table shows the results of a simulation, here for a 4.2 m telescope with 0.6 arc second seeing, photon rate of 10,000 photons per frame and correcting for Zernike modes ranging from NZ=3 (providing tip-tilt correction only) to NZ=15. The table then gives the peak Strehl ratio for each of the selections and the corresponding width of the profile at the levels of levels of 25 % encircled energy, 10 % and 4 %.

At a lower signal level of 100 photons per frame we see in Table 4 that trying to fit 15 Zernike terms produces a marked degradation in the quality of the simulation fit, not only in peak Strehl ratio but at every point throughout the halo. It is clear, however that using up to 10 Zernike terms is still entirely acceptable with the photon rate of 100 photons per frame.

| 4m2 Diameter, 0.6" seeing, 100 Photons per frame, 3/6/10/15 Zernike Modes. | | | | | | | | | | | | | | | | |
|---|---|---|---|---|---|---|---|---|---|---|---|---|---|---|---|---|
| | Strehl ratio | | | | Image diam. in arcsec, at 25 % | | | | Image diam. in arcsec at 10 % | | | | Image diam. in arcsec, at 4 % | | | |
| Selection | NZ 3 | NZ6 | NZ10 | NZ15 | NZ 3 | NZ6 | NZ10 | NZ15 | NZ3 | NZ6 | NZ10 | NZ15 | NZ 3 | NZ6 | NZ10 | NZ15 |
| 10% LI | 0.21 | 0.19 | 018 | 0.14 | 0.31 | 0.36 | 0.37 | 0.50 | 0.68 | 0.73 | 0.77 | 0.98 | 1.05 | 1.13 | 1.18 | 1.46 |
| 50% LI | 0.19 | 0.18 | 0.17 | 0.13 | 0.36 | 0.38 | 0.39 | 0.52 | 0.73 | 0.77 | 0.79 | 1.00 | 1.12 | 1.16 | 1.20 | 1.50 |
| 100% LI | 0.17 | 0.17 | 0.165 | 0.12 | 0.39 | 0.40 | 0.40 | 0.53 | 0.77 | 0.78 | 0.80 | 1.02 | 1.16 | 1.18 | 1.21 | 1.57 |
| Raw | 0.057 | .057 | 0.057 | 0.057 | 0.86 | 0.87 | 0.85 | 0.84 | 1.22 | 1.23 | 1.21 | 1.26 | 1.79 | 1.81 | 1.78 | 2.06 |

**Table 4.** This table shows the results of a simulation, here for a 4.2 m telescope with 0.6 arc second seeing, photon rate of 100 photons per frame and correcting for Zernike modes ranging from NZ=3 (providing tip-tilt correction only) to NZ=15. The table then gives the peak Strehl ratio for each of the selections and the corresponding width of the profile at the levels of 25 % encircled energy, 10 % and 4 %.

Even at the yet lower level of only 30 photons per frame we get reasonably good correction with three Zernike terms and only slightly worse with six. This is shown in Table 5 which makes it clear that a degree of correction is still possible with very low photon rates. We should note, however, that every entry in Table 6 shows a lower value than in Table 4.

That we can work at these photon rates should not come as a surprise. We record images of the out of focus pupil plane on either side of the pupil plane. Tip-tilt manifests itself as a lateral shift of the centre of the images, in the opposite sense on either side of the pupil. Statistically this can be measured fairly accurately even with a small number of photons. Adding in the next set of Zernike terms is still possible but it does degrade the net image quality to a degree.

| 4m2 Diameter, 0.6" seeing, 30 Photons per frame, 3/6/10/15 Zernike Modes. | | | | | | | | | | | | | | | | |
|---|---|---|---|---|---|---|---|---|---|---|---|---|---|---|---|---|
| | Strehl ratio | | | | Image diam. in arcsec, at 25% | | | | Image diam. in arcsec at 10 % | | | | Image diam. in arcsec, at 4 % | | | |
| Selection | NZ 3 | NZ6 | NZ10 | NZ15 | NZ 3 | NZ6 | NZ10 | NZ15 | NZ3 | NZ6 | NZ10 | NZ15 | NZ 3 | NZ6 | NZ10 | NZ15 |
| 10% LI | 0.18 | 0.13 | 0.12 | 0.115 | 0.36 | 0.54 | 0.58 | 0.65 | 0.76 | 1.06 | 1.21 | 1.34 | 1.15 | 1.58 | 1.89 | 2.06 |
| 50% LI | 0.16 | 0.12 | 0.11 | 0.1 | 0.44 | 0.56 | 0.59 | 0.66 | 0.83 | 1.10 | 1.22 | 1.36 | 1.24 | 1.68 | 1.91 | 2.10 |
| 100% LI | 0.14 | 0.11 | 0.10 | 0.09 | 0.47 | 0.57 | 0.60 | 0.67 | 0.88 | 1.13 | 1.25 | 1.37 | 1.33 | 1.74 | 1.98 | 2.13 |
| Raw | 0.057 | .057 | 0.057 | 0.057 | 0.84 | 0.85 | 0.85 | 0.84 | 1.26 | 1.29 | 1.31 | 1.27 | 2.06 | 2.08 | 2.09 | 2.06 |

**Table 5.** This table shows the results of a simulation, here for a 4.2 m telescope with 0.6 arc second seeing, photon rate of 30 photons per frame and correcting for Zernike modes ranging from NZ=3 (providing tip-tilt correction only) to NZ=15. The table then gives the peak Strehl ratio for each of the selections and the corresponding width of the profile at the levels of 25 % encircled energy, 10 % and 4 %.

In Table 6 we can see the effect of atmospheric seeing on the quality of the output images. In this table we have used high photon rates so that lack of photons should not degrade the results. However, we should expect photon rates it to make more difference at lower photon rates when we are working with less good seeing.

Two clear conclusions may be drawn from Table 6. Firstly, for all seeing levels the improvement in Strehl ratio between the raw data and the 100% lucky image data is invariably a factor of ~2.5-3. Secondly, even with poor seeing the improvement in the shape of the halo is improved substantially by this process.

| 4m2 Diameter, 0.6",0.75",1.0",1.5" seeing, 10,000 Photons per frame, 10 Zernike Modes. | | | | | | | | | | | | | | | | |
|---|---|---|---|---|---|---|---|---|---|---|---|---|---|---|---|---|
| | Strehl ratio | | | | Image diam. in arcsec, at 25 % | | | | Image diam. in arcsec at 10 % | | | | Image diam. in arcsec, at 4 % | | | |
| Selection | 0.6" | 0.75" | 1.0" | 1.5" | 0.6" | 0.75" | 1.0" | 1.5" | 0.6" | 0.75" | 1.0" | 1.5" | 0.6" | 0.75" | 1.0" | 1.5" |
| 10% LI | 0.21 | 0.17 | 0.12 | 0.06 | 0.32 | 0.37 | 0.42 | 0.65 | 0.69 | 0.77 | 0.94 | 1.47 | 1.07 | 1.24 | 1.64 | 2.43 |
| 50% LI | 0.18 | 0.14 | 0.10 | 0.06 | 0.38 | 0.41 | 0.48 | 0.69 | 0.74 | 0.82 | 1.02 | 1.55 | 1.14 | 1.35 | 1.78 | 2.53 |
| 100% LI | 0.16 | 0.12 | 0.09 | 0.05 | 0.39 | 0.44 | 0.53 | 0.72 | 0.77 | 0.88 | 1.10 | 1.62 | 1.2 | 1.46 | 1.89 | 2.63 |
| Raw data | 0.057 | 0.046 | 0.034 | 0.019 | 0.840 | 0.91 | 1.090 | 1.770 | 1.270 | 1.660 | 1.930 | 2.710 | 2.060 | 2.340 | 2.800 | 4.130 |

**Table 6**. This table shows the results of a simulation, here for a 4.2 m telescope with seeing in the range of 0.6 – 1.5 arcsecs, photon rate of 10,000 photons per frame and correcting for the first 10 Zernike modes. The table then gives the peak Strehl ratio for each of the selections and the corresponding width of the profile at the levels of 25 % encircled energy, 10 % and 4 %.

In Table 7 we compare the results using camera frame rates from 50 Hz-5 Hz. This shows the effect of using lower performance camera systems or if the user wishes to use particularly faint reference stars by effectively increasing the wavefront sensor exposure time. The effect on the images is very similar to what we see under conditions of hire windspeed. Slow camera rates means that the wavefront is smeared as it moves across the telescope entrance pupil. What we see is that the Strehl ratios, particularly in the 50% and 100% selection bins are significantly lower and the profile of the halo is also significantly broader at the slowest rates. However it is clear that it is still possible to work at low rates. It is worth noting that all the original Lucky Imaging experiments reported by the Cambridge group and summarised by Baldwin et al. (2008) were carried out using an older photon counting camera that ran at 10 Hz.

| 4m2 Diameter, 0.6" seeing, 10,000 Photons per frame, 10 Zernike Modes. Frame rates 50Hz, 25Hz, 10Hz, 5Hz | | | | | | | | | | | | | | | | |
|---|---|---|---|---|---|---|---|---|---|---|---|---|---|---|---|---|
| | Strehl ratio | | | | Image diam. in arcsec, at 25 % | | | | Image diam. in arcsec at 10 % | | | | Image diam. in arcsec, at 4 % | | | |
| Selection | 50 Hz | 25 Hz | 10Hz | 5 Hz | 50 Hz | 25Hz | 10Hz | 5 Hz | 50Hz | 25 Hz | 10 Hz | 5 Hz | 50 Hz | 25 Hz | 10Hz | 5 Hz |
| 10% LI | 0.17 | 0.18 | 0.18 | 0.17 | 0.41 | 0.43 | 0.37 | 0.40 | 0.79 | 0.86 | 0.74 | 0.87 | 1.20 | 1.30 | 1.11 | 1.44 |
| 50% LI | 0.16 | 0.16 | 0.16 | 0.13 | 0.44 | 0.48 | 0.40 | 0.50 | 0.83 | 0.90 | 0.79 | 0.97 | 1.26 | 1.38 | 1.12 | 1.58 |
| 100% LI | 0.14 | 0.14 | 0.13 | 0.11 | 0.46 | 0.50 | 0.43 | 0.56 | 0.87 | 0.94 | 0.82 | 1.05 | 1.33 | 1.46 | 1.26 | 1.70 |
| Raw data | 0.057 | 0.057 | 0.054 | 0.052 | 0.840 | 0.96 | 0.78 | 0.95 | 1.270 | 1.45 | 1.19 | 1.54 | 2.060 | 2.29 | 1.94 | 2.58 |

**Table 7.** This table shows the results of a simulation, here for a 4.2 m telescope with seeing 0.6 arcsec, photon rate of 10,000 photons per frame and correcting for the first 10 Zernike modes. Simulations were done with cameras running at different rates of 50 Hz, 25 Hz, 10 Hz and 5 Hz. The table then gives the peak Strehl ratio for each of the selections and the corresponding width of the profile at the levels of 25 % encircled energy, 10 % and 4 %.

Our conclusions therefore are that with a 4.2 m telescope and 10 Zernike modes we are able to produce a marked improvement on the image quality by combining low order adaptive optics with Lucky Imaging. In many ways the visual perception of the images produced is more important. In Fig 12 we have simulated a random star field with the luminosity function very similar to that of the bulge of our Galaxy in I-band. Each tile shows a region of about 4.2 x 8.3 arc seconds. The left-hand panel shows results obtained with high photon rate of 10,000 photons per frame while the right hand one shows a much lower level of 30 photons per frame. It is surprising how good the processes even with a low photon rate. Close inspection of the images do make it clear that the right-hand panel is significantly poorer particularly at the 50% and 100% selection. The halo on the images with the lower photon rates are much more marked. However in both cases the improvement over the image quality from the raw images is quite marked.

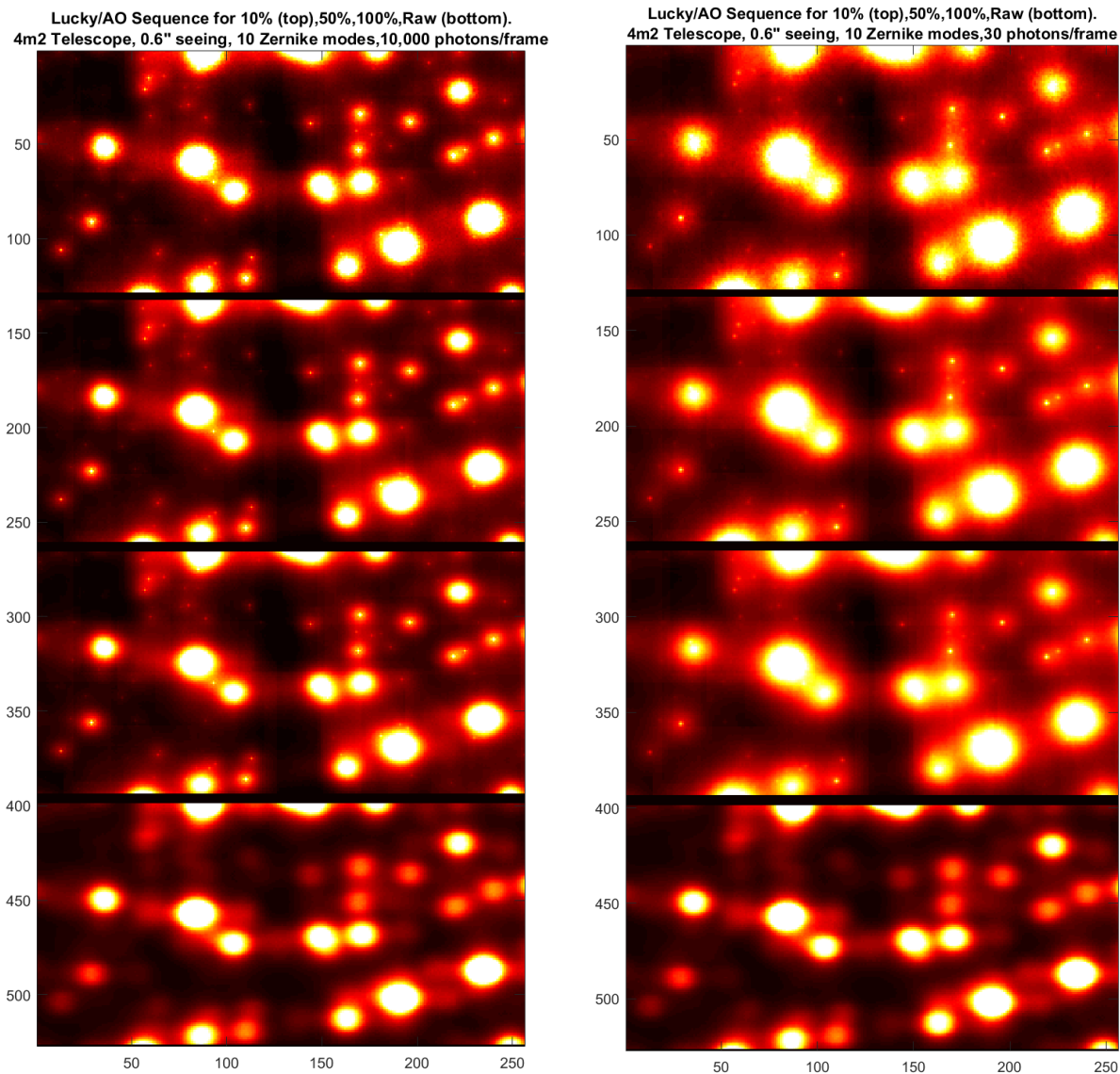

**Fig 12:** a comparison of the simulation results for a 4.2 m telescope, 0.6" seeing, and 10 Zernike mode processing for (a, left-hand panel) signal of 10,000 photons per frame and (b, right-hand panel) signal of only 30 photons per frame. Each subpanel covers an area of about 4.2 x 8.3 arc seconds at 33 milliarcsecs per pixel. The lowest sub image in each panel shows the raw unprocessed data as it would be recorded with a conventional long exposure on that telescope. Above that are sub-images showing the output with 100%, 50% and 10% (top) selection respectively.

### 7.8. 8.2m Telescope Results.

The results from simulations of an 8.2 m telescope are reported here in the same way as for the 4.2 m telescope. For each simulation a set of plots was produced similar to that shown in Fig 11. We ran a series of simulations starting with a high photons per frame number of 10,000 and examining the output images using between 6 and 21 Zernike terms. The results in Table 8 show that the Strehl ratio of the simulated images gradually improves with the use of more Zernike terms. The Strehl ratio for 10% selection is significantly better than it is for 50% and 100%. The Strehl ratio increases only slightly between 6 and 21 Zernike terms. However the size of the halo around the central peak continues to improve up to (and we find beyond) 15 Zernike terms, at every part of the profile of the halo.

| 8m2 Diameter, 0.6" seeing, 10,000 Photons per frame, 6/10/15/21 Zernike Modes. | | | | | | | | | | | | | | | | |
|---|---|---|---|---|---|---|---|---|---|---|---|---|---|---|---|---|
| | Strehl ratio | | | | Image diam. in arcsec at 25% | | | | Image diam. in arcsec at 10 % | | | | Image diam. in arcsec, at 4 % | | | |
| Selection | NZ 6 | NZ10 | NZ15 | NZ21 | NZ 6 | NZ10 | NZ15 | NZ21 | NZ6 | NZ10 | NZ15 | NZ21 | NZ 6 | NZ10 | NZ15 | NZ21 |
| 10% LI | 0.093 | 0.093 | 0.094 | 0.099 | 0.39 | 0.39 | 0.39 | 0.37 | 0.86 | 0.86 | 0.84 | 0.80 | 1.46 | 1.45 | 1.43 | 1.38 |
| 50% LI | 0.086 | 0.087 | 0.089 | 0.093 | 0.42 | 0.41 | 0.40 | 0.38 | 0.92 | 0.89 | 0.86 | 0.83 | 1.54 | 1.51 | 1.47 | 1.42 |
| 100% LI | 0.077 | 0.081 | 0.083 | 0.089 | 0.44 | 0.43 | 0.41 | 0.40 | 0.96 | 0.92 | 0.89 | 0.85 | 1.60 | 1.56 | 1.52 | 1.47 |
| Raw | 0.027 | 0.027 | 0.027 | 0.027 | 1.07 | 1.07 | 1.07 | 1.06 | 1.66 | 1.66 | 1.67 | 1.49 | 2.43 | 2.43 | 2.43 | 2.43 |

**Table 8.** This table shows the results of a simulation, here for a 8.2 m telescope with 0.6 arc second seeing, photon rate of 10,000 photons per frame and correcting for Zernike modes ranging from NZ= 6 to NZ=21. The table then gives the peak Strehl ratio for each of the selections and the corresponding width of the profile at the levels of levels of 25 % encircled energy, 10 % and 4 %.

Repeating these tests but with a much lower photon rate of 100 photons per frame we see in Table 9 very similar performance except that the results with 21 Zernike terms are somewhat poorer than they are with 10 terms. The higher order Zernike terms allow the representation of the wavefront to cover progressively smaller scales in the wavefront. However the fact that the wavefront recovery quality becomes poorer with higher Zernike modes is simply a consequence of their not being enough photons to give a good signal-to-noise over the scales that corresponds to that number of Zernike terms. It is clear, however that using up to 10 Zernike terms is still entirely acceptable with the photon rate of 100 photons per frame.

| 8m2 Diameter, 0.6" seeing, 100 Photons per frame, 6/10/15/21 Zernike Modes. | | | | | | | | | | | | | | | | |
|---|---|---|---|---|---|---|---|---|---|---|---|---|---|---|---|---|
| | Strehl ratio | | | | Image diam. in arcsec at 25% | | | | Image diam. in arcsec at 10 % | | | | Image diam. in arcsec, at 4 % | | | |
| Selection | NZ 6 | NZ10 | NZ15 | NZ21 | NZ 6 | NZ10 | NZ15 | NZ21 | NZ6 | NZ10 | NZ15 | NZ21 | NZ 6 | NZ10 | NZ15 | NZ21 |
| 10% LI | 0.067 | 0.059 | 0.054 | 0.050 | 0.51 | 0.58 | 0.62 | 0.69 | 1.22 | 1.39 | 1.53 | 1.66 | 1.93 | 1.45 | 2.30 | 2.35 |
| 50% LI | 0.063 | 0.056 | 0.053 | 0.049 | 0.54 | 0.59 | 0.60 | 0.66 | 1.26 | 1.42 | 1.54 | 1.66 | 1.98 | 1.51 | 2.31 | 2.37 |
| 100% LI | 0.059 | 0.052 | 0.050 | 0.047 | 0.55 | 0.60 | 0.61 | 0.63 | 1.29 | 1.44 | 1.55 | 1.66 | 2.02 | 1.56 | 2.32 | 2.38 |
| Raw | 0.027 | 0.027 | 0.027 | 0.028 | 1.06 | 1.07 | 1.06 | 1.05 | 1.67 | 1.67 | 1.65 | 1.64 | 2.42 | 2.43 | 2.42 | 2.40 |

**Table 9.** This table shows the results of a simulation, here for a 8.2 m telescope with 0.6 arc second seeing, photon rate of 100 photons per frame and correcting for Zernike modes ranging from NZ=6 to NZ=21. The table then gives the peak Strehl ratio for each of the selections and the corresponding width of the profile at the levels of 25 % encircled energy, 10 % and 4 %.

In Table 10 we can see the effect of atmospheric seeing on the quality of the output images. In this table we have used high photon rates so lack of photons should not affect the results, though we should expect it to make more difference at the lowest photon rates when we are working with less good seeing. Two clear conclusions may be drawn from Table 10. Firstly, for all seeing levels the improvement in Strehl ratio between the raw data and the 100% lucky image data is invariably a factor of ~2.5-3. Secondly, even with poor seeing the improvement in the shape of the halo is improved significantly by this process.

| 8m2 Diameter, 0.6",0.75",1.0",1.5" seeing, 10,000 Photons per frame, 15 Zernike Modes. | | | | | | | | | | | | | | | | |
|---|---|---|---|---|---|---|---|---|---|---|---|---|---|---|---|---|
| | Strehl ratio | | | | Image diam. in arcsec, at 25 % | | | | Image diam. in arcsec at 10 % | | | | Image diam. in arcsec, at 4 % | | | |
| Selection | 0.6" | 0.75" | 1.0" | 1.5" | 0.6" | 0.75" | 1.0" | 1.5" | 0.6" | 0.75" | 1.0" | 1.5" | 0.6" | 0.75" | 1.0" | 1.5" |
| 10% LI | 0.087 | 0.078 | 0.063 | 0.06 | 0.43 | 0.43 | 0.48 | 0.65 | 0.84 | 0.96 | 1.07 | 1.47 | 1.43 | 1.64 | 1.73 | 2.43 |
| 50% LI | 0.080 | 0.072 | 0.056 | 0.06 | 0.46 | 0.47 | 0.51 | 0.69 | 0.86 | 1.02 | 1.15 | 1.55 | 1.47 | 1.71 | 1.80 | 2.53 |
| 100% LI | 0.074 | 0.066 | 0.051 | 0.05 | 0.48 | 0.48 | 0.53 | 0.72 | 0.89 | 1.05 | 1.20 | 1.62 | 1.52 | 1.76 | 1.85 | 2.63 |
| Raw data | 0.027 | 0.022 | 0.016 | 0.019 | 1.04 | 1.27 | 1.56 | 1.770 | 1.67 | 2.02 | 2.26 | 2.710 | 2.43 | 2.80 | 3.23 | 4.130 |

**Table 10.** This table shows the results of a simulation, here for a 8.2 m telescope with seeing in the range of 0.6 – 1.5 arcsec, a photon rate of 10,000 photons per frame and correcting for the first 15 Zernike modes. The table then gives the peak Strehl ratio for each of the selections and the corresponding width of the profile at the levels of 25 % encircled energy, 10 % and 4 %.

Our conclusions therefore are that with a 8.2 m telescope and 15 Zernike modes we are able to produce a marked improvement on the image quality by combining low order adaptive optics with LI. In many ways the visual perception of the images produced is more important. In Fig 13 we have simulated a random star field with the luminosity function very similar to that of the bulge of our Galaxy in I-band. Each tile shows a region of about 5 x 10 arc seconds. The left-hand image shows results obtained with high photon rate of 10,000 photons per frame while the right hand one shows a much lower level of 30 photons per frame. It is surprising how good the process is even with a low photon rate. Close inspection of the images do make it clear that the right-hand panel is significantly poorer particularly at the 50% and 100% selection. The halo on the images with the lower photon rate are much more marked. However in both cases the improvement over the image quality from the raw images is quite marked.

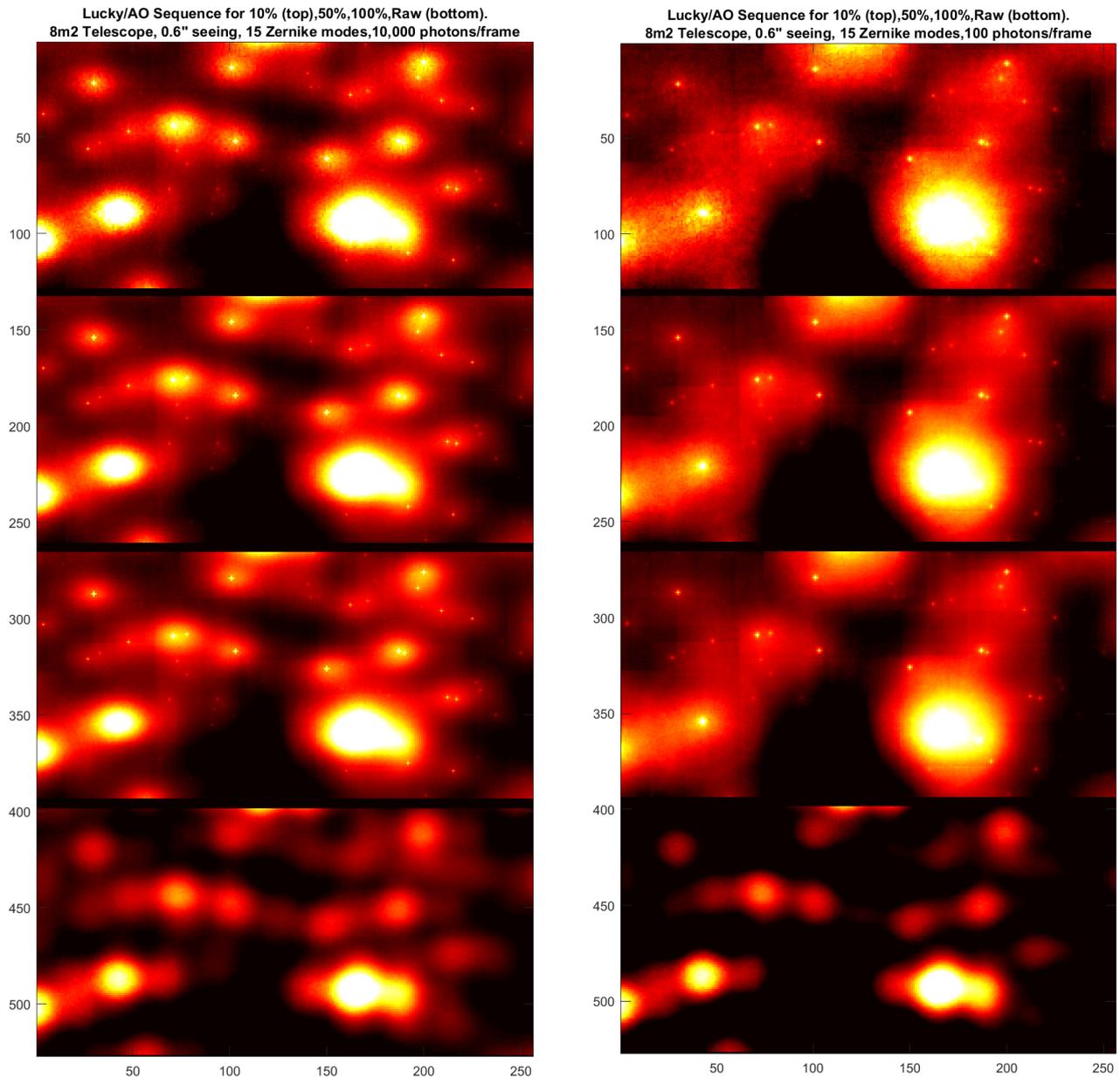

**Fig 13**: a comparison of the simulation results for a 8.2 m telescope, 0.6" seeing, and 15 Zernike mode processing for (a, left-hand panel) signal of 10,000 photons per frame and (b, right-hand panel) signal of only 100 photons per frame. Each subpanel covers an area of about 2.5 x 5 arc seconds at 20 milliarcsecs per pixel. The lowest sub image in each panel shows the raw unprocessed data as it would be recorded with a conventional long exposure on that telescope. Above that are some images showing the output with 100%, 50% and 10% (top) selection respectively.

### 7.9. Simulations With A 2.5M Telescope

Although the main drive for this study of the improvement of telescope point spread functions with lower order AO has been the consequences for telescopes in the 4-8 m class, the same methods may be applied to smaller telescopes with considerable success. We have already seen (Fig 1, for example) that a 2.5 m telescope on a good site can routinely produce images with a resolution very close to that achieved by the Hubble Space Telescope, simply by removing tip tilt (the essence of the LI technique). The same procedures that we have used above show that if we correct more than 3 Zernike terms beyond tip-tilt to correct for defocus and some astigmatism the images are also further improved.

The results in Table 11 show that the Strehl ratio of the simulated images gradually improves with the use of more Zernike terms. The Strehl ratio for 10% selection is significantly better than it is for 50% and 100%. However the size of the halo around the central peak continues to improve up to (and we find beyond) 15 Zernike terms, at every part of the profile of the halo.

| 2m5 Diameter, 0.6" seeing, 10,000 Photons per frame, 3/6/10/15 Zernike Modes. | | | | | | | | | | | | | | | | |
|---|---|---|---|---|---|---|---|---|---|---|---|---|---|---|---|---|
| | Strehl ratio | | | | Image diam. in arcsec, at 25 % | | | | Image diam. in arcsec at 10 % | | | | Image diam. in arcsec, at 4 % | | | |
| Selection | NZ 3 | NZ6 | NZ10 | NZ15 | NZ 3 | NZ6 | NZ10 | NZ15 | NZ3 | NZ6 | NZ10 | NZ15 | NZ 3 | NZ6 | NZ10 | NZ15 |
| 10% LI | 0.34 | 0.36 | 0.42 | 0.47 | 0.29 | 0.27 | 0.20 | 0.14 | 0.68 | 0.63 | 0.55 | 0.49 | 1.09 | 1.03 | 0.94 | 0.86 |
| 50% LI | 0.28 | 0.32 | 0.38 | 0.42 | 0.35 | 0.34 | 0.27 | 0.16 | 0.77 | 0.70 | 0.58 | 0.52 | 1.19 | 1.08 | 0.99 | 0.90 |
| 100% LI | 0.26 | 0.30 | 0.36 | 0.40 | 0.42 | 0.35 | 0.29 | 0.18 | 0.82 | 0.70 | 0.63 | 0.54 | 1.24 | 1.12 | 1.01 | 0.93 |
| Raw | 0.095 | 0.095 | 0.095 | 0.095 | 0.86 | 0.87 | 0.87 | 0.86 | 1.16 | 1.27 | 1.28 | 1.28 | 1.92 | 1.92 | 1.92 | 1.90 |

**Table 11**. This table shows the results of a simulation, here for a 2.5 m telescope with 0.6 arc second seeing, photon rate of 10,000 photons per frame and correcting for Zernike modes ranging from NZ=3 (providing tip-tilt correction only) to NZ=15. The table then gives the peak Strehl ratio for each of the selections and the corresponding width of the profile at the levels of levels of 25 % encircled energy, 10 % and 4 %.

Repeating these tests but with a much lower photon rate of 100 photons per frame we see in Table 12 very similar performance except that the results with 15 Zernike terms are markedly poorer than they are with 10 terms. The higher order Zernike terms allow the representation of the wavefront to cover progressively smaller scales in the wavefront.

At a yet lower signal level of 100 photons per frame we see in Table 12 that trying to fit 15 Zernike terms produces a marked degradation in the quality of the simulation fit, not only in peak Strehl ratio but at every point throughout the halo. It is clear, however that using up to 10 Zernike terms is still entirely acceptable with the photon rate of 100 photons per frame.

| 2m5 Diameter, 0.6" seeing, 100 Photons per frame, 3/6/10/15 Zernike Modes. | | | | | | | | | | | | | | | | |
|---|---|---|---|---|---|---|---|---|---|---|---|---|---|---|---|---|
| | Strehl ratio | | | | Image diam. in arcsec, at 25 % | | | | Image diam. in arcsec at 10 % | | | | Image diam. in arcsec, at 4 % | | | |
| Selection | NZ 3 | NZ6 | NZ10 | NZ15 | NZ 3 | NZ6 | NZ10 | NZ15 | NZ3 | NZ6 | NZ10 | NZ15 | NZ 3 | NZ6 | NZ10 | NZ15 |
| 10% LI | 0.34 | 0.34 | 0.34 | 0.36 | 0.30 | 0.32 | 0.31 | 0.27 | 0.67 | 0.66 | 0.66 | 0.63 | 1.07 | 1.07 | 1.07 | 1.03 |
| 50% LI | 0.29 | 0.31 | 0.32 | 0.33 | 0.38 | 0.34 | 0.33 | 0.30 | 0.76 | 0.71 | 0.69 | 0.65 | 1.17 | 1.14 | 1.10 | 1.08 |
| 100% LI | 0.26 | 0.28 | 0.30 | 0.32 | 0.43 | 0.39 | 0.35 | 0.33 | 0.82 | 0.76 | 0.71 | 0.69 | 1.24 | 1.18 | 1.14 | 1.10 |
| Raw | 0.095 | 0.095 | 0.095 | 0.095 | 0.89 | 0.88 | 0.88 | 0.88 | 1.29 | 1.23 | 1.28 | 1.28 | 1.96 | 1.93 | 1.92 | 1.94 |

**Table 12**. This table shows the results of a simulation, here for a 2.5 m telescope with 0.6 arc second seeing, photon rate of 100 photons per frame and correcting for Zernike modes ranging from NZ=3 (providing tip-tilt correction only) to NZ=15. The table then gives the peak Strehl ratio for each of the selections and the corresponding width of the profile at the levels of 25 % encircled energy, 10 % and 4 %.

Even at the yet lower level of only 10 photons per frame we get good correction with three Zernike terms and only slightly worse with six. This is shown in Table 13 and it makes it clear that a degree of correction is still possible with remarkably low photon rates. That we can work at these photon rates should not surprise us. We record images of the out of focus pupil plane on either side of the pupil plane. Tip-tilt manifests itself as a lateral shift of the centre of the images, in the opposite sense on either side of the pupil. Statistically this can be measured fairly accurately even with a small number of photons. Adding in the next set of Zernike terms is still possible but it does degrade the net image quality to a degree.

| 2m5 Diameter, 0.6" seeing, 10 Photons per frame, 3/6/10/15 Zernike Modes. | | | | | | | | | | | | | | | | |
|---|---|---|---|---|---|---|---|---|---|---|---|---|---|---|---|---|
| | Strehl ratio | | | | Image diam. in arcsec, at 25 % | | | | Image diam. in arcsec at 10 % | | | | Image diam. in arcsec, at 4 % | | | |
| Selection | NZ 3 | NZ6 | NZ10 | NZ15 | NZ 3 | NZ6 | NZ10 | NZ15 | NZ3 | NZ6 | NZ10 | NZ15 | NZ 3 | NZ6 | NZ10 | NZ15 |
| 10% LI | 0.32 | 0.25 | 0.23 | 0.2 2 | 0.33 | 0.44 | 0.48 | 0.49 | 0.71 | 0.90 | 0.98 | 1.03 | 1.11 | 1.38 | 1.50 | 1.58 |
| 50% LI | 0.28 | 0.24 | 0.215 | 0.20 | 0.39 | 0.46 | 0.50 | 0.54 | 0.77 | 0.92 | 1.02 | 1.10 | 1.19 | 1.42 | 1.56 | 1.65 |
| 100% LI | 0.25 | 0.23 | 0.205 | 0.195 | 0.43 | 0.47 | 0.51 | 0.54 | 0.83 | 0.95 | 1.03 | 1.10 | 1.25 | 1.45 | 1.58 | 1.66 |
| Raw | 0.095 | 0.095 | 0.095 | 0.095 | 0.88 | 0.88 | 0.88 | 0.88 | 1.28 | 1.28 | 1.28 | 1.28 | 1.92 | 1.94 | 1.92 | 1.93 |

**Table 13**. This table shows the results of a simulation, here for a 2.5 m telescope with 0.6 arc second seeing, photon rate of 30 photons per frame and correcting for Zernike modes ranging from NZ=3 (providing tip-tilt correction only) to NZ=15. The table then gives the peak Strehl ratio for each of the selections and the corresponding width of the profile at the levels of 25 % encircled energy, 10 % and 4 %.

In Table 14 we can see the effect of atmospheric seeing on the quality of the output images. In this table we have used high photon rates so that should not affect the outcome, though we should expect it to make more difference at the lowest photon rates when we are working with less good seeing. Two clear conclusions may be drawn from Table 14. Firstly, for all seeing levels the improvement in Strehl ratio between the raw data and the 100% lucky image data is invariably a factor of ~2.5-3. Secondly, even with poor seeing the improvement in the shape of the halo is improved substantially by this process.

| 2m5 Diameter, 0.6",0.75",1.0",1.5" seeing, 10,000 Photons per frame, 6 Zernike Modes. | | | | | | | | | | | | | | | | |
|---|---|---|---|---|---|---|---|---|---|---|---|---|---|---|---|---|
| | Strehl ratio | | | | Image diam. in arcsec, at 25 % | | | | Image diam. in arcsec at 10 % | | | | Image diam. in arcsec, at 4 % | | | |
| Selection | 0.6" | 0.75" | 1.0" | 1.5" | 0.6" | 0.75" | 1.0" | 1.5" | 0.6" | 0.75" | 1.0" | 1.5" | 0.6" | 0.75" | 1.0" | 1.5" |
| 10% LI | 0.36 | 0.29 | 0.22 | 0.15 | 0.28 | 0.34 | 0.40 | 0.85 | 0.64 | 0.74 | 0.92 | 1.94 | 1.03 | 1.20 | 1.48 | 3.13 |
| 50% LI | 0.32 | 0.26 | 0.185 | 0.13 | 0.33 | 0.38 | 0.51 | 1.00 | 0.69 | 0.79 | 1.03 | 2.09 | 1.08 | 1.26 | 1.60 | 3.37 |
| 100% LI | 0.30 | 0.24 | 0.16 | 0.12 | 0.34 | 0.42 | 0.55 | 1.11 | 0.70 | 0.83 | 1.09 | 2.23 | 1.13 | 1.31 | 1.69 | 3.56 |
| Raw data | 0.094 | 0.77 | 0.061 | 0.019 | 0.88 | 0.99 | 1.14 | 2.33 | 1.28 | 1.46 | 1.74 | 3.66 | 1.93 | 2.20 | 2.59 | 5.42 |

**Table 14**. This table shows the results of a simulation, here for a 2.5 m telescope with seeing in the range of 0.6 – 1.5 arcsecs, photon rate of 10,000 photons per frame and correcting for the first 10 Zernike modes. The table then gives the peak Strehl ratio for each of the selections and the corresponding width of the profile at the levels of 25 % encircled energy, 10 % and 4 %.

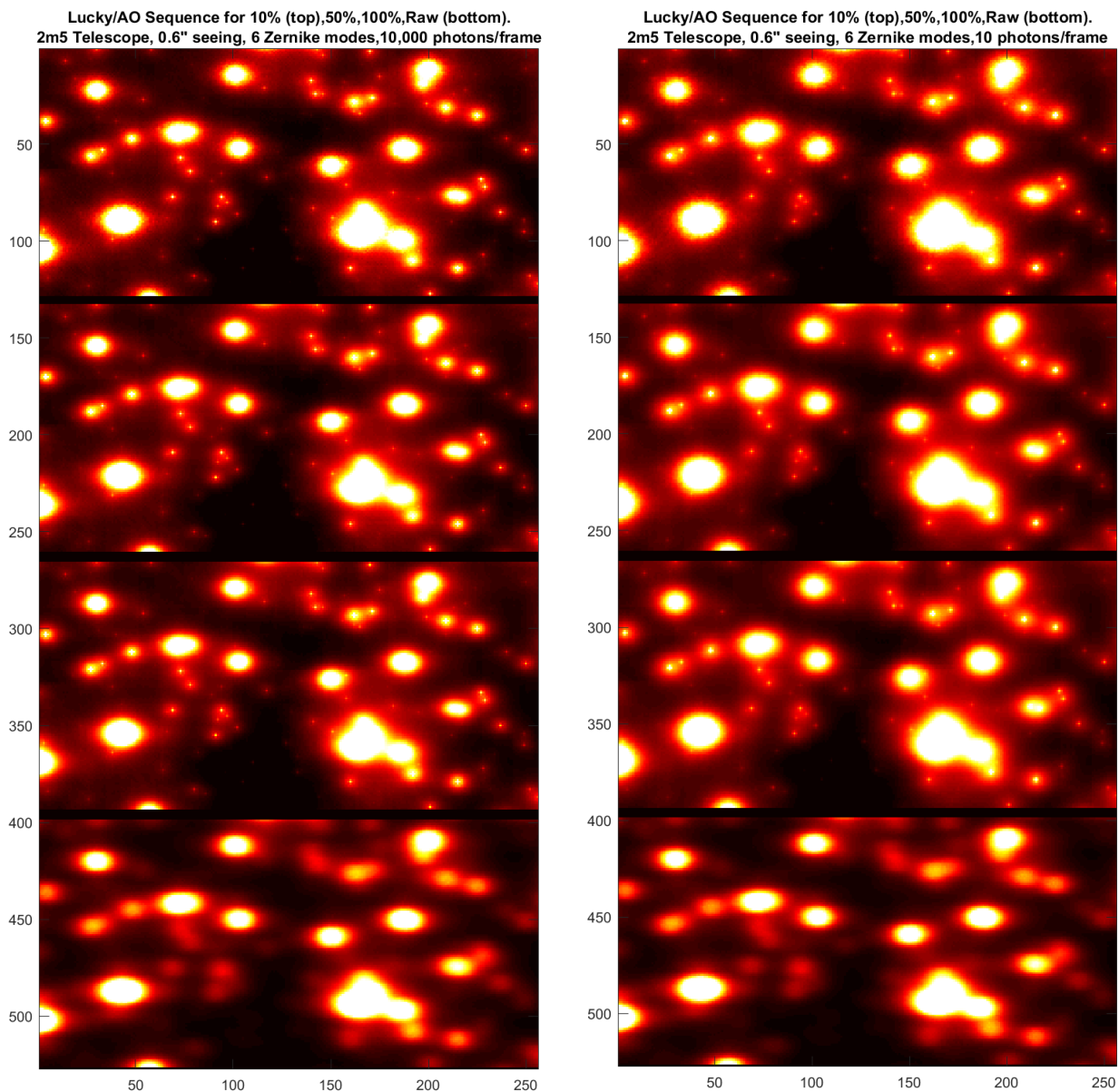

**Fig 14**: a comparison of the simulation results for a 2.5 m telescope, 0.6" seeing, and 6 Zernike mode processing for (a, left-hand panel) signal of 10,000 photons per frame and (b, right-hand panel) signal of only 10 photons per frame. Each subpanel covers an area of about 5.8 x 11.5 arc seconds at 45 milliarcsecs per pixel. The lowest sub image in each panel shows the raw unprocessed data as it would be recorded with a conventional long exposure on that telescope. Above that are some images showing the output with 100%, 50% and 10% (top) selection respectively.

We have a considerable amount of experience of using LI techniques on 2.5 m telescopes although all were without a low order AO system to complement them. The results that were obtained when only tip tilt correction was simulated match very closely those that were obtained on the 2.5 m NOT telescope under very similar seeing conditions. These have been described in detail by Baldwin, Warner & Mackay (2008). The results shown in figure 14 may be compared directly with those shown in Fig 1.

### 7.10. 3.6m Telescope with Lucky Imaging alone.

A recent proposal for a new instrument called GravityCam has been developed that is described in detail by Mackay et al. (2018). It is an instrument planned to survey a 0.5° field on the New Technology Telescope (NTT) of the European Southern Observatory at La Silla in Chile. Simulations of the performance of the telescope have been carried out using exactly the same principles described above. In this case it will depend purely on LI and therefore the simulations correspond to only using the first three Zernike modes. The simulated performance is described in the following Tables.

In Table 15, the performance is simulated using the same procedures described above. As before we see that the raw data Strehl ratio is improved under all seeing conditions by a factor in the range of 2.5-3. There are significant improvements to be made even with poor seeing. The site of the NTT in Chile has a median seeing of 0.75 arc seconds and it is clear that at the high signal level of 1000 photons per frame the correction is excellent.

| 3.6 Diameter, 0.6",0.75",1.0",1.5" seeing, 1000 Photons per frame, 3 Zernike Modes. | | | | | | | | | | | | | | | | |
|---|---|---|---|---|---|---|---|---|---|---|---|---|---|---|---|---|
| | Strehl ratio | | | | Image diam. in arcsec, at 25 % | | | | Image diam. in arcsec at 10 % | | | | Image diam. in arcsec, at 4 % | | | |
| Selection | 0.6" | 0.75" | 1.0" | 1.5" | 0.6" | 0.75" | 1.0" | 1.5" | 0.6" | 0.75" | 1.0" | 1.5" | 0.6" | 0.75" | 1.0" | 1.5" |
| 10% LI | 0.215 | 0.17 | 0.13 | 0.08 | 0.39 | 0.47 | 0.63 | 0.87 | 0.82 | 1.01 | 1.48 | 2.04 | 1.32 | 1.67 | 2.44 | 3.39 |
| 50% LI | 0.18 | 0.145 | 0.115 | 0.07 | 0.45 | 0.54 | 0.78 | 1.02 | 0.92 | 1.12 | 1.67 | 2.21 | 1.45 | 1.79 | 2.64 | 3.60 |
| 100% LI | 0.16 | 0.13 | 0.096 | 0.63 | 0.52 | 0.60 | 0.84 | 1.10 | 0.98 | 1.19 | 1.78 | 2.34 | 1.55 | 1.88 | 2.79 | 3.81 |
| Raw data | 0.065 | 0.052 | 0.04 | 0.024 | 0.97 | 1.17 | 1.77 | 2.36 | 1.54 | 1.85 | 2.67 | 3.77 | 2.24 | 2.75 | 4.16 | 5.69 |

**Table 15**. This table shows the results of a simulation, here for a 3.6 m telescope with seeing in the range of 0.6 – 1.5 arcsec, photon rate of 1,000 photons per frame and correcting for the first 3 (tip -tilt) Zernike modes only. The table then gives the peak Strehl ratio for each of the selections and the corresponding width of the profile at the levels of 25 % encircled energy, 10 % and 4 %.

At lower signal levels of 100 photons per frame (Table 16) and 10 photons per frame (Table 17) the performance is still good although at the lowest photon rates and poorer seeing the performance is inevitably affected.

| 3m6 Diameter, 0.6",0.75",1.0",1.5" seeing, 100 Photons per frame, 3 Zernike Modes. | | | | | | | | | | | | | | | | |
|---|---|---|---|---|---|---|---|---|---|---|---|---|---|---|---|---|
| | Strehl ratio | | | | Image diam. in arcsec, at 25 % | | | | Image diam. in arcsec at 10 % | | | | Image diam. in arcsec, at 4 % | | | |
| Selection | 0.6" | 0.75" | 1.0" | 1.5" | 0.6" | 0.75" | 1.0" | 1.5" | 0.6" | 0.75" | 1.0" | 1.5" | 0.6" | 0.75" | 1.0" | 1.5" |
| 10% LI | 0.215 | 0.17 | 0.125 | 0.084 | 0.39 | 0.45 | 0.66 | 0.84 | 0.82 | 0.99 | 1.46 | 2.04 | 1.32 | 1.64 | 2.36 | 3.34 |
| 50% LI | 0.18 | 0.145 | 0.107 | 0.072 | 0.45 | 0.54 | 0.76 | 1.00 | 0.92 | 1.13 | 1.62 | 2.23 | 1.45 | 1.80 | 2.54 | 3.63 |
| 100% LI | 0.16 | 0.13 | 0.095 | 0.062 | 0.52 | 0.60 | 0.81 | 1.10 | 0.98 | 1.21 | 1.69 | 2.36 | 1.55 | 1.91 | 2.68 | 3.86 |
| Raw data | 0.094 | 0.052 | 0.038 | 0.025 | 0.97 | 1.19 | 1.74 | 2.38 | 1.54 | 1.88 | 2.61 | 3.81 | 2.24 | 2.80 | 4.08 | 5.78 |

**Table 16**. This table shows the results of a simulation, here for a 3.6 m telescope with seeing in the range of 0.6 – 1.5 arcsec, photon rate of 100 photons per frame and correcting for the first 3 (tip -tilt) Zernike modes only. The table then gives the peak Strehl ratio for each of the selections and the corresponding width of the profile at the levels of 25 % encircled energy, 10 % and 4 %.

| 3m6 Diameter, 0.6",0.75",1.0",1.5" seeing, 10 Photons per frame, 3 Zernike Modes. | | | | | | | | | | | | | | | | |
|---|---|---|---|---|---|---|---|---|---|---|---|---|---|---|---|---|
| | Strehl ratio | | | | Image diam. in arcsec, at 25 % | | | | Image diam. in arcsec at 10 % | | | | Image diam. in arcsec, at 4 % | | | |
| Selection | 0.6" | 0.75" | 1.0" | 1.5" | 0.6" | 0.75" | 1.0" | 1.5" | 0.6" | 0.75" | 1.0" | 1.5" | 0.6" | 0.75" | 1.0" | 1.5" |
| 10% LI | 0.22 | 0.17 | 0.126 | 0.08 | 0.34 | 0.47 | 0.69 | 0.85 | 0.75 | 1.00 | 1.55 | 1.94 | 1.21 | 1.20 | 2.52 | 3.13 |
| 50% LI | 0.18 | 0.146 | 0.108 | 0.07 | 0.42 | 0.55 | 0.81 | 1.00 | 0.85 | 1.11 | 1.72 | 2.09 | 1.34 | 1.26 | 2.71 | 3.37 |
| 100% LI | 0.16 | 0.128 | 0.095 | 0.062 | 0.48 | 0.60 | 0.87 | 1.11 | 0.91 | 1.20 | 1.82 | 2.23 | 1.44 | 1.31 | 2.86 | 3.56 |
| Raw data | 0.066 | 0.050 | 0.039 | 0.025 | 0.90 | 1.19 | 1.85 | 2.33 | 1.43 | 1.87 | 2.77 | 3.66 | 2.07 | 2.20 | 4.33 | 5.42 |

**Table 17** This table shows the results of a simulation, here for a 3.6 m telescope with seeing in the range of 0.6 – 1.5 arcsec, photon rate of 100 photons per frame and correcting for the first 3 (tip -tilt) Zernike modes only. The table then gives the peak Strehl ratio for each of the selections and the corresponding width of the profile at the levels of 25 % encircled energy, 10 % and 4 %.

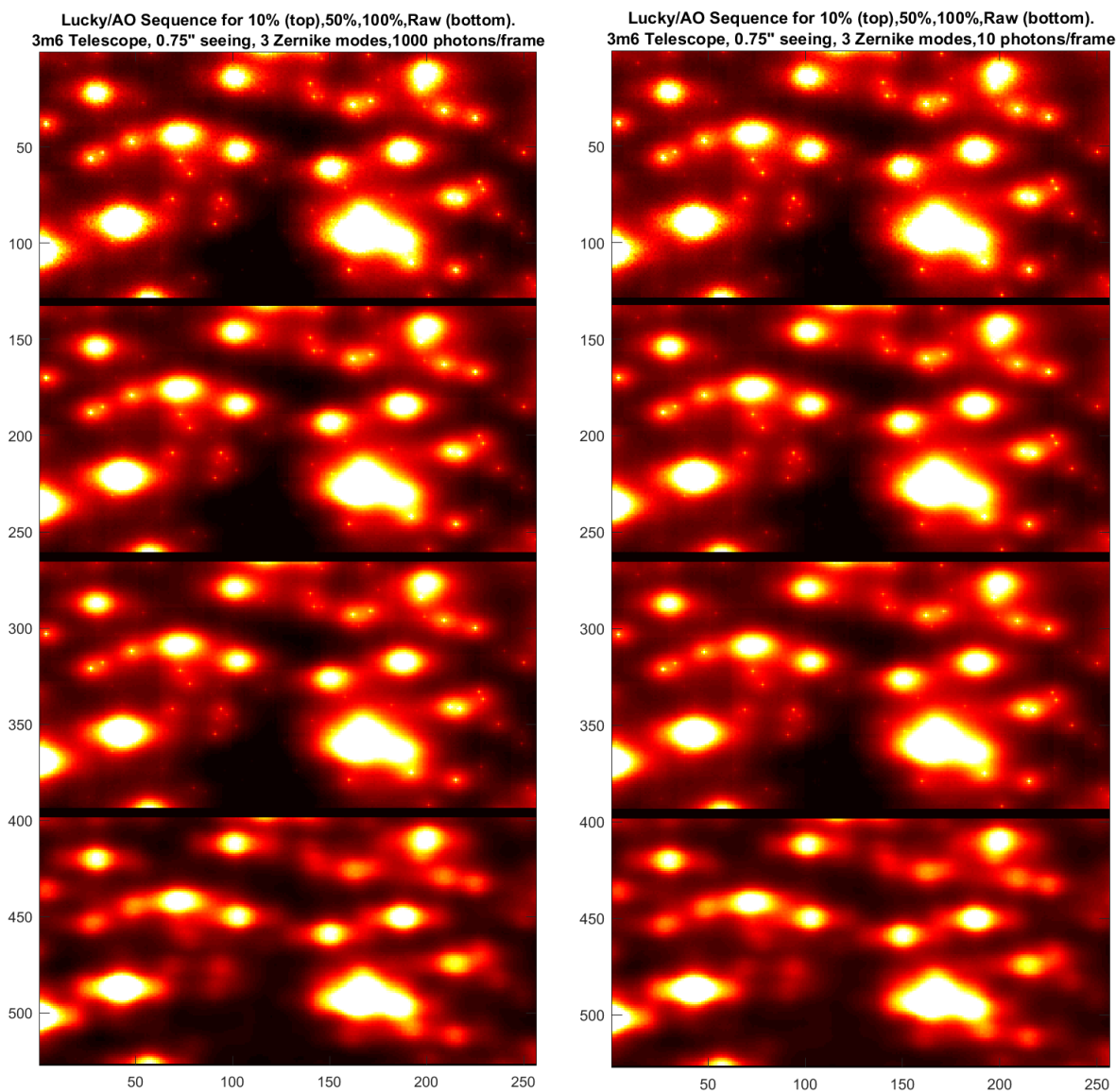

**Fig 15**: a comparison of the simulation results for a 3.6 m telescope, 0.75" seeing the median seeing on the NTT telescope where the GravityCam would be mounted), and 3 Zernike mode processing for (a, left-hand panel) signal of 1000 photons per frame and (b, right-hand panel) signal of only 10 photons per frame. Each subpanel covers an area of about 6.1 x 12.3 arc seconds at 48 milliarcsecs per pixel. The lowest sub image in each panel shows the raw unprocessed data as it would be recorded with a conventional long exposure on that telescope. Above that are some images showing the output with 100%, 50% and 10% (top) selection respectively.

One of the key requirements for all the instruments using LI either with or without AO and including GravityCam is their capacity to work over wide fields of view. The correction processes with LI either with or without adaptive optics are less able to provide good improvement as the angular distance between the target field and the science field increases. In crowded fields this can be managed by using many different reference stars to cover areas much smaller than the field of even a single detector which makes cover many minutes of arc. In less crowded fields care must be taken in choosing reference objects so that the effects of anisoplanatism are minimised.

## 8. Technical Discussion and Implications.

### 8.1. Anisoplanatism.

The discussion in this paper so far has centred almost exclusively on the properties of the reference star image to be used for both LI on its own and in conjunction with low order AO. It is critically important for the success of the technique that the properties of images relatively nearby the reference star should be recognisably similar to those of the star. We say that the target objects need to be within the region of isoplanatism of the reference star. The light entering the pupil of the telescope from a target star offset from the reference star will pass through a different part of the atmosphere so we should expect in principle that the image profile match between the two will not be perfect.

This is discussed in section 3.2 of Baldwin, Warner & Mackay (2008). Our simulation package allows this to be checked. The wavefront error is assessed for the reference star using however many Zernike terms used in the particular simulation under consideration. We calculate the phase errors in the wavefront that has passed through an offset part of the turbulent screen. That is then corrected using the phase errors derived for the reference object. This slightly less well corrected wavefront allows the offset image profile to be computed and checked for Strehl ratio relative to that achieved for the reference star. The isoplanatic patch size is defined as the diameter of the area within which the images show a Strehl ratio reduced by a factor of $1/e$ (2.72). An example of the results obtained is shown in Fig 16.

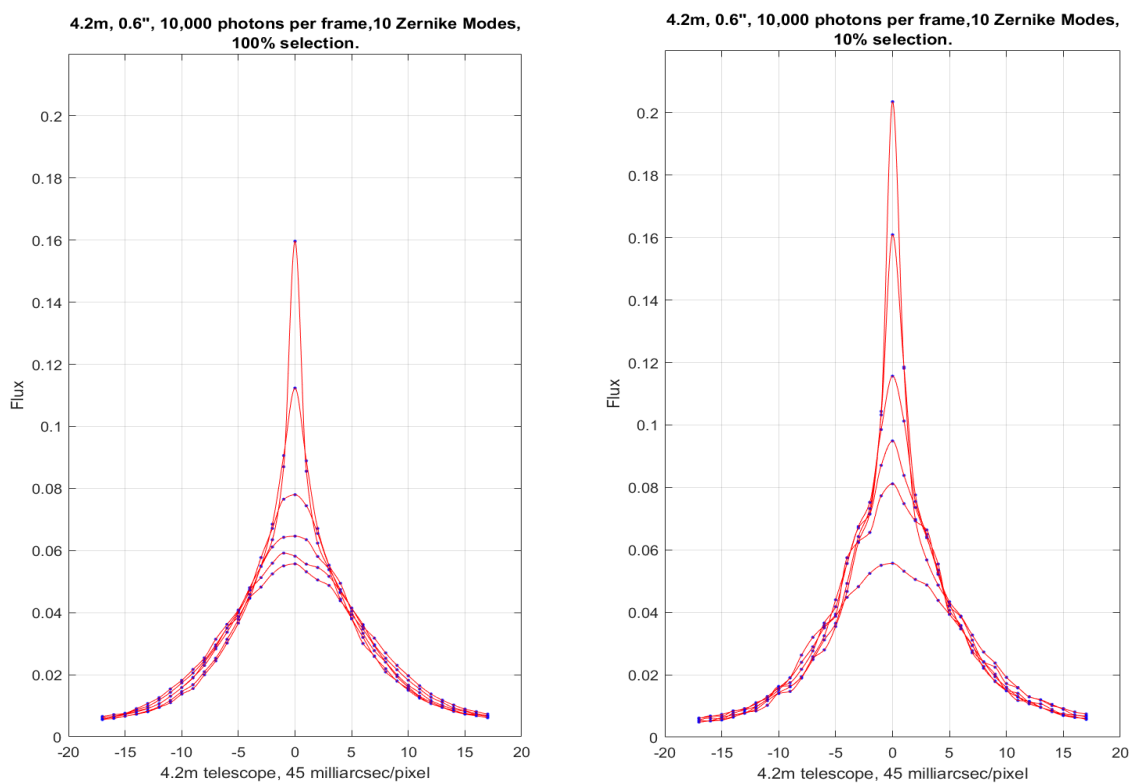

**Fig 16:** sets of one-dimensional profiles through images offset relative to the reference star. If we assume that a typical altitude for the turbulence we are dealing with is ~10 km then each of the profiles corresponds to an additional offset of about ~10 arcsec from the reference star. The top curve shows the profile for the reference star itself. The left-hand plot with simulations of 0.6 arc second seeing and 100% selection show an isoplanatic patch radius of 30 arcsec. The right hand simulation shows the results with the same 0.6 arcsec second seeing but now with 10% selection shows a similar or slightly larger isoplanatic patch radius. We would predict that this effect is purely a geometric effect set by the size of the telescope pupil and the altitude of the turbulent layer. We do not find the isoplanatic patch size to be greatly affected by the particular level of seeing, or by the fraction of images selected to synthesise the lucky output image. These calculations suggest that an isoplanatic patch in the region of one arc minute diameter should be achievable.

In practice we have found quite frequently that the isoplanatic patch size is significantly larger than implied by the simulations. It is most likely due to the high altitude turbulence being relatively weak with most turbulence being generated at much lower levels in the atmosphere, and often referred to as ground-layer turbulence. This would be predicted to give much larger isoplanatic patch sizes and therefore allow a wider choice of reference object even at high galactic latitudes.

**8.2. Reference Star Constraints.**

The results presented above show that images may be combined to give a much sharper image using reference stars that produce only a few tens or hundreds of photons per frame time. All these calculations have assumed a 25 Hz frame rate although we know from the results presented by Baldwin, Warner & Mackay (2008), and table 7 that we can work successfully with significantly slower frame rates. The limit will be set by the wind speed at the time observing. Slower frame rates risk smearing the wavefront in the direction of the wind. We have worked successfully with frame rates as low 8 Hz giving a sensitivity improvement three-fold compared to the magnitudes given in many of the above figures.

Further improvements are possible with careful bandpass filtering. The bandpass for the science field does not need to be the same as the bandpass for the reference star. In addition, there is no need to restrict the filter bandpass to those used for conventional photometry. If we increase the bandpass the phase patterns we measure become the sum of the phase patterns at each wavelength. Essentially the phase pattern expands as we use longer wavelengths but, even with a relatively broad pass band, and we know that the turbulence scales as $\sim \lambda^{-1.2}$. The effect on the derived phase pattern with a broader bandpass is relatively small and still a reasonably good representation of what we would record with a narrowband.

There have been significant improvements in the performance of silicon detectors in recent years, both charge-coupled devices and CMOS detectors. Deep depletion technologies (see Mackay et al, 2019) allow the far red response to be improved substantially, increasing the potential of these devices for working at longer wavelengths. Quantum efficiencies in excess of 90% are now routine at wavelengths longer than 900 nm. By using a reference star band set simply by a long pass filter allowing light with wavelength $\lambda$ in the range 700 - 1100nm would improve the system sensitivity by a further factor of three.

| Photon rates/frame vs, telescope size and filter/CCD combinations. | | | | | | |
|---|---|---|---|---|---|---|
| | 1000 photons/frame | | 100 photons/frame | | 10 photons/frame | |
| Telescope | standard | enhanced | standard | enhanced | standard | enhanced |
| 2.5 m | 14.9 | 16.3 | 16.4 | 17.8 | 18.9 | 20.3 |
| 3.6 m | 15.5 | 16.9 | 17 | 18.4 | 19.5 | 20.9 |
| 4.2 m | 16 | 17.4 | 17.5 | 18.9 | 20 | 21.4 |
| 8.2 m | 17.5 | 18.9 | 19 | 20.4 | 21.5 | 22.9 |

**Table 18:** the above simulations have all been carried out with simply a specified number of detected photons per frame. The detected photon rates listed in the second row of the table and assuming 25 Hz frame rate and remembering that we detect to separate near-pupil planes are then shown for each of the four telescope sizes discussed in the paper correspond to limiting I band magnitudes for an M2-type star when we use a standard thinned CCD/CMOS detector and an I pass band filter for the reference star and wavefront measuring system (in the columns marked "standard"). By using a deep depletion CCD/CMOS detector such as are now being produced by TeledyneE2V fainter limiting magnitudes may be achieved and are shown in the columns marked "enhanced".

At high galactic latitudes reference stars become more and more difficult to find. At I ~ 20, there is about one star per square arc minute with galaxies surface densities several times higher. It is often imagined that one can only use a star as a reference object but galaxies, provided they are compact enough may also be used. At lower galactic latitudes there are many more stars making the choice of reference object significantly more convenient and, importantly, closer to the science target under study.

It must be appreciated that the use of a broad filter pass band to increase the photon detection rate from the reference star does not mean that the same broad band must be used for the remainder of the field. Once the phase errors have been compensated for properly, those corrections will apply throughout the field at least within the isoplanatic patch constraints. The science field may be restricted to a much narrower band, for example a rest wavelength narrowband H-alpha filter could still be used on nearby targets. Inevitably, wavefront correction carried out from far red measurements will be poorer when applied to a much shorter wavelength of light but not so much poorer as to be useless. The instrumental configuration for a low order AO plus LI system might separate the wavefront sensing path from the science path. This is what has been done in the AOLI instrument described below.

The conclusion should be then that with appropriate instrument design reference objects much fainter than for any other design of wavefront sensor may well be usable for LI when combined with low order AO.

**8.3 Spectroscopic Applications.**

This paper has addressed the ways that images with substantially improved angular resolution may be delivered with relatively simple and relatively inexpensive solutions. The reader should not get the impression, however, that these methodologies are only relevant to direct imaging. Improved angular resolution is just as important for spectroscopic applications as they are for imaging. The techniques described here allow for example a LI/AO instrument to feed efficiently a multi-object fibre-fed spectrographs. The constraints set by the finite isoplanatic patch sizes cannot be ignored, however. The image quality will degrade with angular distance from the bright reference object and so science object and reference object cannot be too far apart for ideal correction.

The total light from each fibre can be measured to establish the location of the centroid or brightest speckle in the image. This will guide the alignment of consecutive frames. With fast readout detectors running at typical LI rates we would see the individual spectral channels fluctuating in integrated brightness. Selecting the channel with the highest signal level allows the registration of all the individual spectra to be identified before being summed. This would allow the image quality delivered by conventional LI to significantly improve fibre bundle fed spectroscopic science.

When LI is combined with low order adaptive optics the instrumental setup is the same. The corrected images are fed onto a fibre bundle and the processing done in a very similar way to that described elsewhere in the paper.

**9. Conclusions**

The atmospheric turbulence that compromises astronomical imaging even on a good site has most of its power concentrated in the lowest orders, corresponding to the largest scales across the telescope aperture. We have simulated the effects of eliminating largely the lowest turbulence orders on astronomical images and shown that with a novel design of a curvature wavefront sensor we can achieve sensitivities that allow much of the atmospheric turbulence to be eliminated even with very faint reference stars. These techniques offer for the first time the prospect of being able to undertake significantly higher resolution imaging studies of the sky with natural guide stars with I band magnitudes in the range of $16^m$-$20^m$. Observations combining lucky imaging and a large aperture telescope have already been published showing near diffraction-limited performance under good seeing with a 5 m class telescope.

Our approach to the design of a curvature wavefront sensor has several important advantages over the limitations imposed by Shack-Hartmann wavefront sensors. Firstly the light level needed from a reference star does not have to be achieved in each and every lenslet of the array. The turbulent structure across the pupil is large-scale and it is only necessary to have enough light within a corresponding fraction of the total telescope mirror area. In turn it means that the readout speed is set by the decorrelation that might be expected over one of these larger areas and not across the size of one single lenslet, a size that is very much smaller and which therefore requires much faster system frame rate. The technique is also very stable at the lowest photon rates and will only fit those Zernike terms that are appropriate for that rate.

These techniques can be applied equally to observations in the near-IR. There are already near-IR detectors available which are capable of fast read-out rates. The read noise of these detectors is now very good although we generally find the photon rates available in the IR are much greater than in the visible. The readout speed requirement is less severe in the IR and the phase errors are reduced proportional to ~ $\lambda^{-1.2}$. This again makes it easier to compensate for phase variance introduced by atmospheric turbulence.

It is clear that these techniques have a great potential for improving the images delivered by ground-based telescopes. It is important to recognise that there is much more that can be done to further improve the resolution of astronomical images affected by atmospheric turbulence. For example the work of Schodel et al (2013) and Bosco et al (2019) uses holographic techniques to further improve not simply resolution but the improvement of the isoplanatic patch size in these images.

It is inevitable that the Hubble Space Telescope will eventually stop working. The James Webb Space Telescope has only limited capacity in the visible, if indeed it launches successfully. Inevitably it will be left to the instruments used on ground-based telescopes that will have to deliver science of the highest quality. We believe that the techniques described here will make a significant contribution to what can be achieved from the ground.

## 10. Acknowledgements

We are very grateful to Marcos van Dam and collaborators for providing the outline code used for our simulations. It proved to be very efficiently and thoughtfully written. We benefited from helpful discussions with Alex Oscoz, Sergio Velasco, Juan Valdivia, Carlos Colondro-Conde, Marta Puga and others at the IAC, Tenerife. The Nordic Optical Telescope is operated on the island of La Palma jointly by Denmark, Finland, Iceland, Norway and Sweden in the Spanish Observatorio del Roque de los Muchachos of the IAC. We are grateful to the efforts of the PALMAO team of the Palomar Observatory for their assistance through our run on the 5 m telescope. The author is also grateful to the Institute of Astronomy of the University of Cambridge for providing facilities for this research.

The author is particularly grateful to an anonymous referee whose comprehensive suggestions about presentation and style have helped to improve this paper greatly. Finally, the author would particularly like to acknowledge the help and support over many years of Prof John Baldwin who sadly died on 2010. A good friend, and an exceptional colleague who is still greatly missed.

**REFERENCES**

Baldwin, J.E., Beckett, M.G., Boysen, R.C., Burns, D., Buscher, D.F., Cox, G.C., Haniff, C.A., Mackay, C.D., Nightingale, N.S., Rogers, J., Scheuer, P.A.G., Scott, T.R., Tuthill, P.G., Warner, P.J., Wilson, D.M.A., Wilson, R.W., ( 1996), A&A, 306, L13-L16.

Baldwin, J.E, Warner, P.J., and Mackay, C.D., (2008), A&A, vol. 480, p589.

Basden, A. G., Haniff, C. A., and Mackay, C. D.,(2003), MNRAS, vol. 345, 985-991.

Bennett, D. P., and Rhie, S. H., (1996) Ap. J. vol 472, 2

Bosco, F., Pott, J.-U., and Shodel, R., (2019), PASP, vol 131, DOI:10.1088/1538-3873/ab019f

Crass, J., King,D., Mackay, C., (2014), SPIE 9148-81, Montréal, June 2014, doi:10.1117/12.925714

Fried, D. L.(1967), Proc. IEEE, Volume 55, p. 57-67, 57–67

Fried, D L,(1978), JOSA, vol 68, p1651-1658.

Graves, J. E., Northcott, M. J.,Roddier, F. J., Roddier, C. A. and   Close, L. M., (1998) , in Adaptive Optical Systems Technologies, D. Bonaccini and R. K. Tyson, eds., Proc. SPIE    3353, 34-43.

Guyon, O., (2010), PASP, vol 122, 49-62.

Guyon, O., et al, (2008),PASP, 120, 655.

Harding,C.M., Johnston, R. A., & Lane, R.G. (1999), Applied   Optics, 38, 11.

Hardy J. W., , ed. 1998, "Adaptive optics for astronomical telescopes, Adaptive optics for astronomical telescopes"

Hecquet, J., Coupinot, G., (1984), J Optics, vol 15, p375-383.

Hufnagel, R.E., (1966), in "Restoration of Atmospheric Degraded Images", National Academy Of Sciences, Washington, DC, vol 3, Appendix 2, p11.

Johnston, R. A. & Lane, R. G., (2000),Applied Optics, 39, 26.

Johnston, R. A., (2000), PhD thesis "inverse problems in   astronomical imaging", University of Canterbury, New Zealand, http://hdl.handle.net/10092/12858 .

Kaiser, N., Tonry, J L, and Luppino, G.A., (2000), PASP, 112,  p768.

Law, N. M.,Mackay, C. D.,Dekany, R. G.,Ireland, M.,Lloyd, J. P., Moore, A. M., Robertson, J. G.,Tuthill, Woodruff, H., (2009), ApJ **692** p.924-930

Mackay, C.D., ARA&A, 24, 255, 1986.

Mackay, C, et al. (2012b),  Proc SPIE 8446, Amsterdam, doi:10.1117/12.925618

Mackay, C., Weller, K., and Suess, F., (2012a), SPIE vol 8453-1, 2012.

Mackay,C., Basden,A., Bridgeland, M. (2004),"Astronomical imaging with L3CCDs: detector performance and high-speed controller design", SPIE 5499 Glasgow, June 2004, 203-209.

Mackay, C., et al (2016) , SPIE 9908-21. DOI: 10.1117/12.2230900

Mackay, C. et al (2018) "GravityCam: Wide-Field High-Resolution High-Cadence Imaging Surveys in the Visible from the  Ground",PASA, 35, 47. DOI: 10.1017/pasa.2018.43.

Mackay, C. et al, (2019), "Towards High-Resolution Astronomical Imaging", *Astronomy & Geophysics*, Vol 60, Issue 3,June  2019, Pages 3.22–3.27,https://doi.org/10.1093/astrogeo/atz146.

Marin, E., et al, (2018), Proc SPIE 10703, DOI: 10.1117/12.2312768

Murphy, D. V, (1992), The Lincoln Laboratory Journal, 5, 24.
Noll, R. J., (1976), JOSA, vol 66, p207-211.
Pozzetti, L., et al., (1998),MNRAS, vol 298, p1133-1144.
Racine, R., (2006), PASP, 118, 1066.
Racine. R., (1996), "The Telescope Point-Spread Function, PASP, vol 108, 699.
Roddier, F., (1988), Applied Optics, vol. 27, p1223-1225.
Sarazin, M., and Tokovinin, A.,(2001) "The Statistics of Isoplanatic Angle and Adaptive Optics Time Constant Derived from DIMM Data", Venice 2001 conference proceedings on "Beyond Conventional Adaptive Optics".
Schodel, R., et al., (2013), MNRAS vol 429,p1367-1375.
Simons, D., (1995), " Longitudinally Averaged R-Band Field Star Counts across the Entire Sky ", Gemini Observatory technical note.
Sivaramakrishnan, A., et al, (2001)" Ground-based Coronagraphy with High-order Adaptive Optics"Ap.J., vol 552, 1.
Starck, J.L., Donoho, D. L., and Candes, E. J., (2003). A&A 398, 785. DOI: 10.1051/0004-6361:20021571
Tatarski, W. I., (1961), "Wave Propagation in a Turbulent Medium" (McGraw-Hill, New York, 1961).
Trujillo, I. Aguerri, J. A. L. ,Cepa,, J. And Gutiérrez,, C. M. (2001), MNRAS, 328, 977
van Dam, M. A., and Lane, R.G., (2002), Applied Optics, vol.41, p5497.
Velasco, S., et al, (2016), MNRAS, 460, 3519, DOI: 10.1093/mnras/stw1071